\documentclass[a4paper,fleqn,usenatbib]{mnras}

\usepackage{xspace}
\usepackage{amsmath}
\usepackage{mathrsfs}
\usepackage{graphicx}
\usepackage{mathptmx}
\usepackage{txfonts}
\usepackage{ae,aecompl}
\usepackage{url}

\def\xte{\textsl{RXTE}\xspace}

\def\inte{\textsl{INTEGRAL}\xspace}
\def\nustar{\textsl{NuSTAR}\xspace}
\def\swift{\textsl{Swift}\xspace}
\newcommand{\xmm}{\textsl{XMM-Newton}\xspace}
\newcommand{\pn}{\text{EPIC-pn}\xspace}
\newcommand{\mos}{\text{EPIC-MOS}\xspace}
\newcommand{\rgs}{\text{RGS}\xspace}

\def\b12{E_{\rm 12}}

\def\igr{IGR~J00291+5934\xspace}
\def\sax{SAX~J1808.4$-$3658\xspace}
\usepackage{color}

\title[A 8\,mHz QPO from IGR~J00291+5934]{Discovery of a soft X-ray 8\,mHz QPO from the accreting millisecond pulsar \igr}

\author[C. Ferrigno et al. ]{C. Ferrigno$^{1}$\thanks{E-mail:
    carlo.ferrigno@unige.ch (CF)}, E. Bozzo $^{1}$, A. Sanna$^{2}$, F. Pintore$^{3}$, A. Papitto$^{4}$, A. Riggio$^{2}$,\newauthor 
    L. Burderi$^{2}$, T. Di Salvo$^{5}$, R. Iaria$^{5}$, A. D'A\`i$^{6}$
\\ 
$^{1}$ISDC, Department of Astronomy, University of Geneva, Chemin d'Ecogia 16, CH-1290 Versoix, Switzerland\\
$^{2}$Dipartimento di Fisica, Universit\`a degli Studi di Cagliari, SP Monserrato-Sestu km 0.7, I-09042 Monserrato, Italy\\
$^{3}$INAF-Istituto di Astrofisica Spaziale e Fisica Cosmica - Milano, via E. Bassini 15, I-20133 Milano, Italy\\
$^{4}$INAF, Osservatorio Astronomico di Roma, Via di Frascati 33, I-00044, Monteporzio Catone (Roma), Italy\\
$^{5}$Universit\`a degli Studi di Palermo, Dipartimento di Fisica e Chimica, via Archirafi 36, I-90123 Palermo, Italy\\
$^{6}$INAF/IASF Palermo, via Ugo La Malfa 153, I-90146, Palermo, Italy
}

\date{Received ---; accepted ---}

\pubyear{2016}

\begin{document}
\label{firstpage}
\pagerange{\pageref{firstpage}--\pageref{lastpage}}
\maketitle

\begin{abstract}
In this paper, we report on the analysis of the peculiar X-ray variability displayed by the accreting millisecond 
X-ray pulsar \igr in a 80\,ks-long joint \nustar and \xmm observation performed during the source outburst in 2015. 
The light curve of the source was characterized by a flaring-like behavior, with 
typical rise and decay time scales of $\sim$120~s. The flares are accompanied by a remarkable spectral variability, with the X-ray emission 
being generally softer at the peak of the flares. A strong quasi periodic oscillation (QPO) is detected at $\sim$8\,mHz in the power 
spectrum of the source and clearly 
associated with the flaring-like behavior. This feature has the strongest power 
at soft X-rays ($\la$3\,keV). We carried out a dedicated hardness-ratio resolved spectral analysis and a QPO phase-resolved 
spectral analysis, together with an in-depth study of the source timing properties, to investigate 
the origin of this behavior. We suggest that the unusual variability of \igr  observed by \xmm and \nustar 
could be produced by an heartbeat-like mechanism, similar to that operating in black-hole X-ray binaries. 
The possibility that this variability, and the associated QPO, are triggered by phases of quasi-stable nuclear burning, as suggested 
in the literature for a number of other neutron star binaries displaying a similar behavior, 
cannot be solidly tested in the case of \igr due to the paucity of type-I X-ray bursts observed from this source.
\end{abstract}


\section{Introduction}
\label{sec:intro}

\igr is one of the known accreting millisecond X-ray pulsars (AMXPs), i.e. low mass X-ray binaries (LMXBs) 
hosting  a rapidly spinning neutron star and an evolved low mass companion. These systems are characterized by orbital 
periods ranging from 40 minutes to a few hours and spend most of their lifetime in a quiescent state, with a typical 
X-ray luminosity of 10$^{31}$-10$^{32}$~erg~s$^{-1}$. They undergo sporadic X-ray outbursts 
lasting from a few weeks to months, during which the accretion onto the compact object is largely increased 
and the resulting X-ray luminosity can be as high as $\sim$10$^{36}$-10$^{38}$~erg~s$^{-1}$  
\citep[see, e.g.][for a recent review]{patruno2012}.  

\igr is known to host the fastest rotating neutron star among the AMXPs, with a spin period of 1.67~ms. 
The source was discovered for the first time during an outburst in 2004 
\citep{markwardt2004,falanga2005} and displayed a double-peaked outburst in 2008 \citep{lewis2010,patruno2010,hartman2011}. 
Archival \xte\ data revealed that two other outbursts from \igr\ might have occurred in 1998 and 2001 \citep{remillard2004}. 
\begin{figure}
\begin{center}
\includegraphics[angle=0,width=8.0cm]{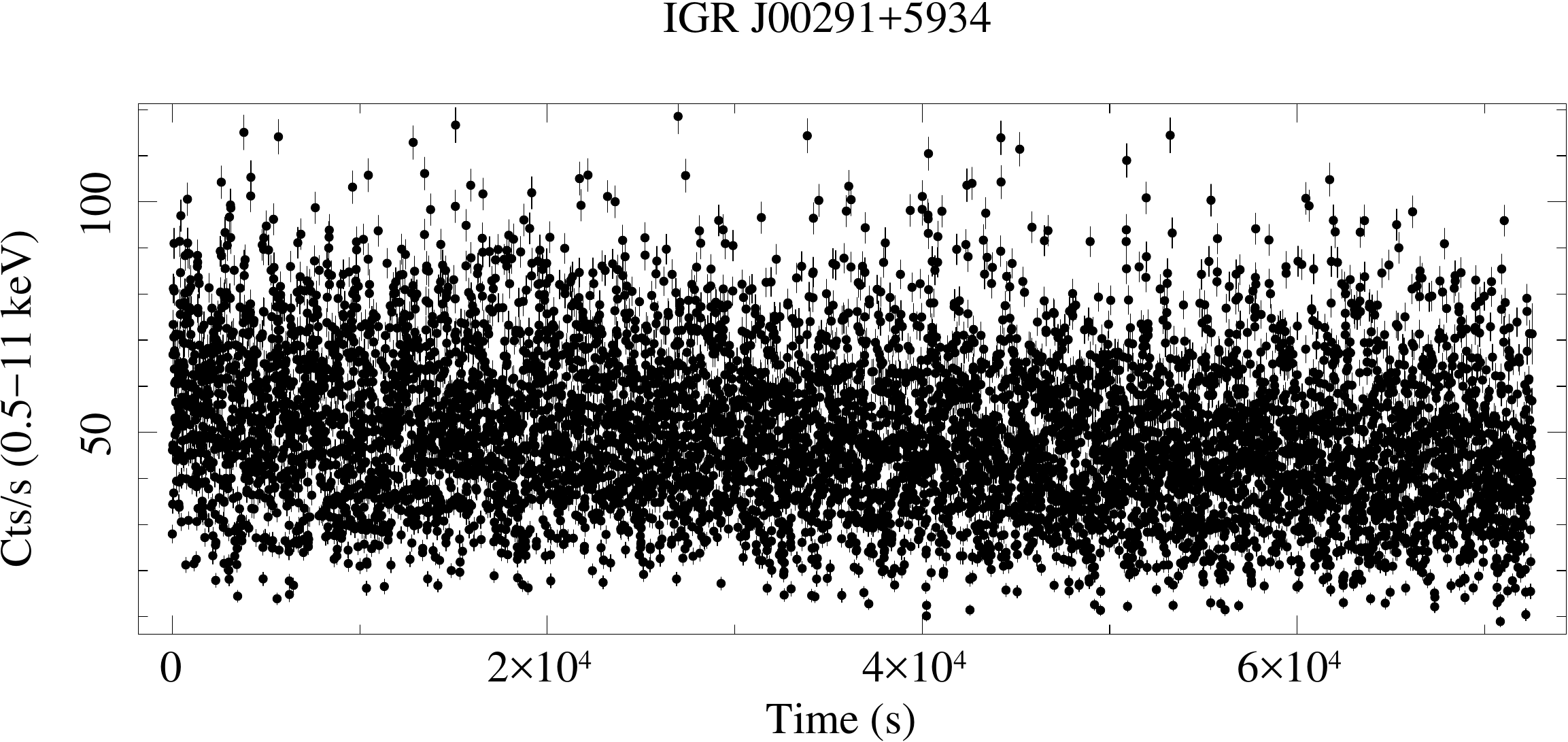}
\includegraphics[angle=0,width=8.0cm]{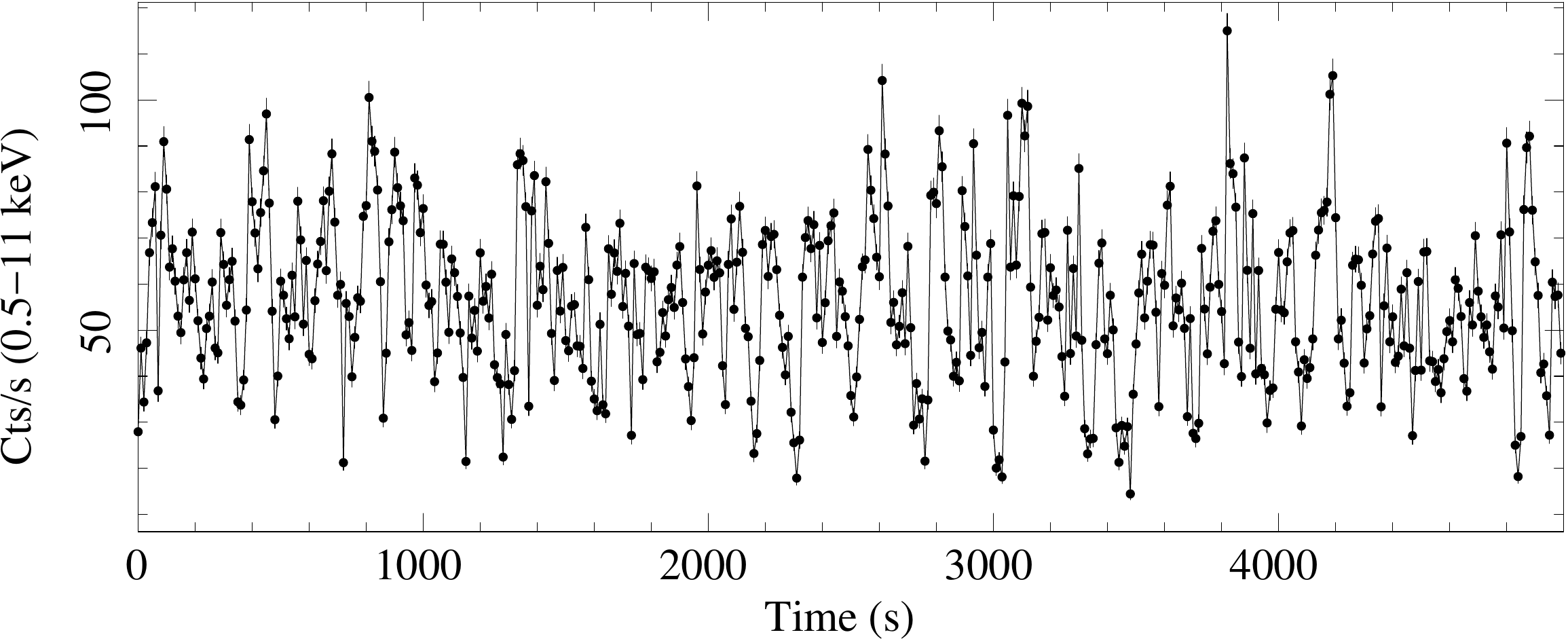}
\caption{{\it Upper panel}: background subtracted light curve of \igr\ as observed by the \pn. The time bin of the light curve is 
10~s and the start time is 2015 July 28 at 12:41 (UT). {\it Lower panel}: a zoom into the first 5\,ks of data with the same time bin 
to highlight better the source variability.}
\label{fig:lcurvepn}
\end{center}
\end{figure}

The latest outburst from \igr was detected in 2015 \citep{sanna2015} and closely monitored with \swift and \inte 
\citep[][, hereafter DF16]{DeFalco2016}. A simultaneous \xmm and \nustar observation of the source was also carried out during this 
event in order to study its broad-band spectral properties. The results obtained from the complete analysis of the \xmm and 
\nustar data has been reported in a complementary publication from our group (Sanna et al. 2016, submitted; hereafter S16). 
In the present paper, we focus on the peculiar variability that characterizes the 
soft X-ray emission of the source during the 2015 outburst. We investigate the aperiodic variability in the
\xmm data and discuss mechanisms that could give rise to the observed phenomenology. 
We complement our report with some results obtained from a simultaneous \nustar observation.

\section{Observations and data analysis}
\label{sec:observations}

During the \xmm observation of \igr, carried out on 2015 July 28 (OBSID 0790181401, see also S16), 
the \pn and the {\mos}2 were operated in {\it TIMING} mode, while the {\mos}1 was operated in small window. To analyze these data, 
we followed the standard data reduction procedures using the 
{\sc epchain} and {\sc emchain} tasks included in SAS v.15.0. 
No episodes of enhanced solar 
activity were revealed in the data, and thus the full exposure time available for all EPIC cameras was retained for 
scientific analysis. The count rate of the source was low enough not to cause any significant pile-up in the \pn and the 
{\mos}2. However, the {\mos}1 was largely affected by pile-up issues and thus discarded for further analysis. 
We extracted the \pn spectra and light curves of the source (background) from the CCD 
columns RAWX=[34--43] ([3--5]). For the extraction of the MOS2 products, we used all events detected from the source 
within the RAWX columns 293--320 of the chip operated in {\it TIMING} mode. The background products were obtained from the chips operating 
in {\it imaging} mode selecting a source-free region. We verified {\it a posteriori} that different reasonable choices of the 
background extraction region did not affect significantly the obtained results.  
We restricted our analysis of the full resolution EPIC spectra to the energy interval 
2.3--11\,keV for the \pn and 0.5--10.0\,keV for the {\mos}2 in order to limit the impact of any instrumental energy calibration 
uncertainty \citep[see also][and references therein]{ferrigno2014}. This restriction was not necessary for the coarsely rebinned spectra 
used for the analysis presented at the end of Sect.~\ref{sec:timing}. 
For completeness, we also report in the sections below some results obtained from the \nustar observation 
carried out simultaneously with \xmm (OBSID.~90101010002). We extracted for both instruments on-board \nustar
(FPMA and FPMB) cleaned event files for the source by performing the same standard screening and filtering 
discussed by S16. We refer the reader to this paper for additional \nustar data analysis details that are 
omitted here for brevity.

\section{The X-ray variability and spectral changes}
\label{sec:spectra}

The light curve of the source as observed by the \pn is reported in Fig.~\ref{fig:lcurvepn} with a bin time of 
10\,s. It can be easily noticed that the source displayed a peculiar variability in the X-ray domain, with quasi-periodic flares 
raising and fainting down on a time scale of roughly 100\,s. 

\begin{figure}
\begin{center}
\includegraphics[angle=0,width=7.8cm]{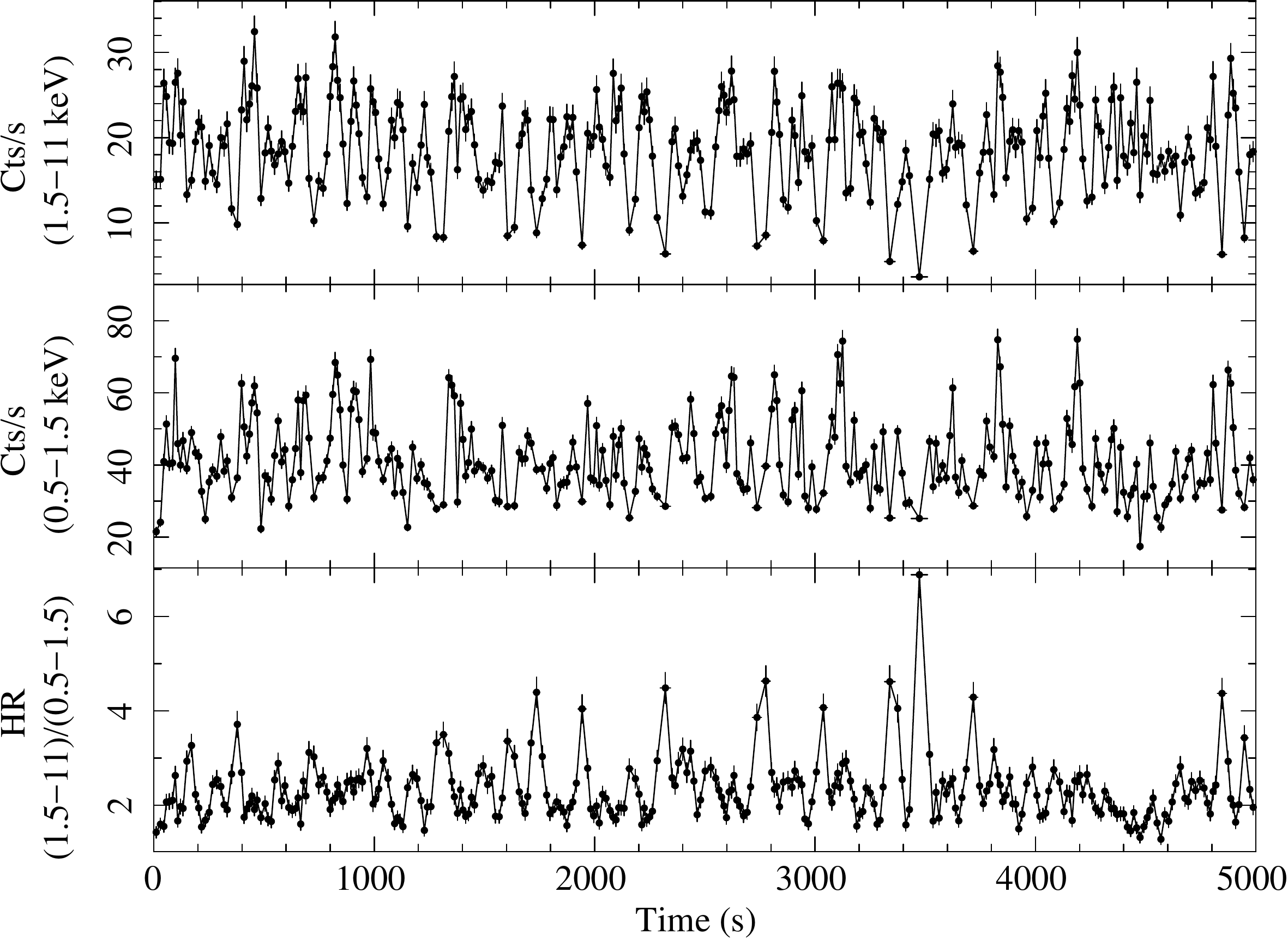}
\caption{A zoom into a representative part of the energy resolved light curves of \igr\ and the correspondingly computed 
hardness ratio, calculated as the ratio between the source count rate in the hard versus soft energy band. 
An adaptive binning has been used in order to achieve in each time bin a minimum S/N of 15 and a minimum bin size of 10\,s;
the start time is 2015 July 28 at 12:41 (UT)}
\label{fig:lc}
\end{center}
\end{figure}

\begin{figure}
\begin{center}
\includegraphics[angle=0,width=7.8cm]{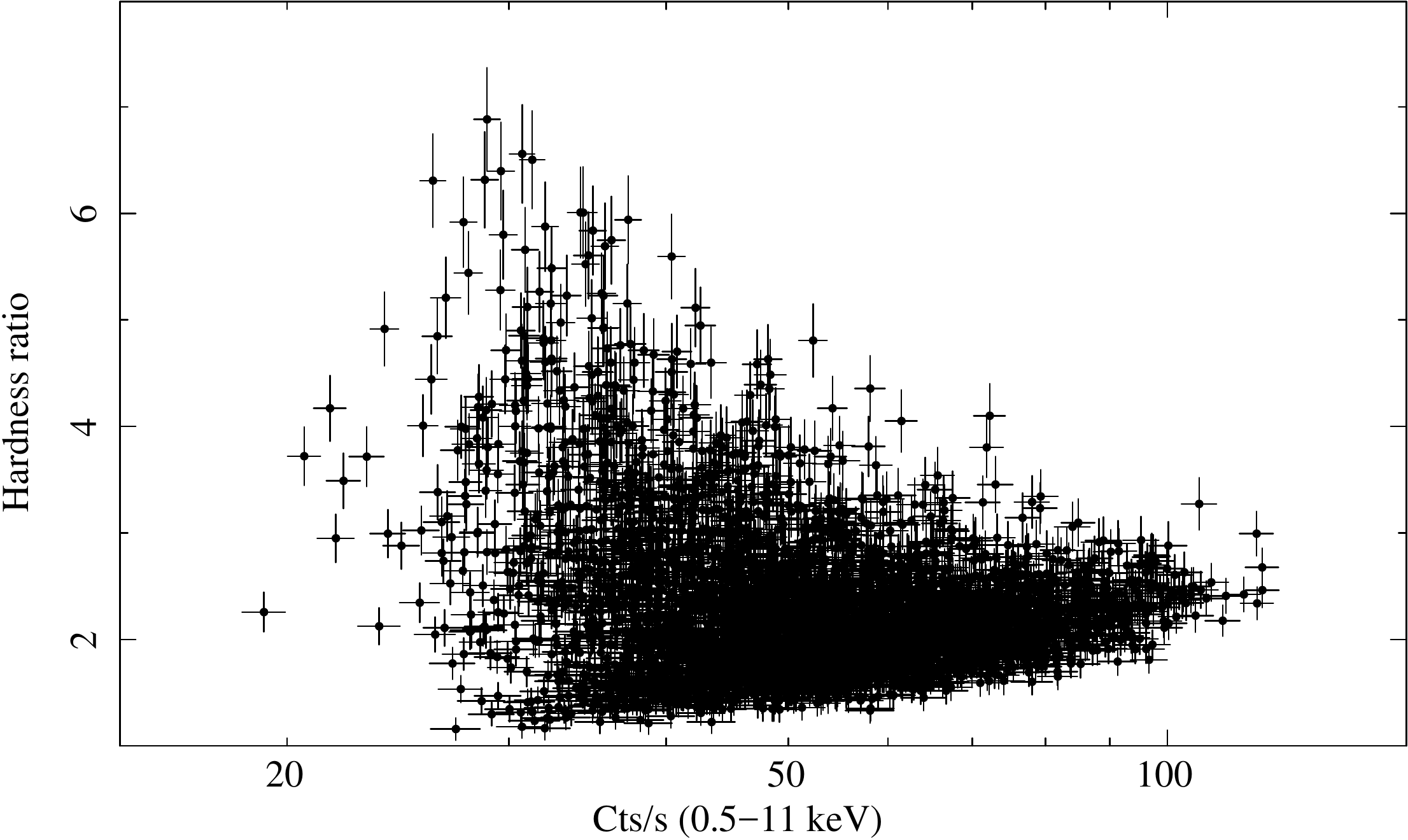}
\caption{Hardness intensity diagram, extracted from the light curves of Fig.~\ref{fig:lc}. The source intensity is 
calculated in the 0.5--11\,keV energy band.}
\label{fig:HID}
\end{center}
\end{figure}

To investigate the origin of this variability, we first computed the ratio between the light curves 
in the 1.5--11\,keV and 0.5--1.5\,keV energy ranges. We used an adaptive rebinning in order to achieve  
a signal-to-noise ratio (S/N) of at least 15 in each time bin and maintain a minimum bin duration of 10\,s. 
The result is shown in Fig.~\ref{fig:lc}. It is evident from this figure that   
the emission of the source becomes generally (but not always) harder between one flare and the other, with the   
hardness ratio (HR) inversely changing with respect to the trend of the source intensity. 
This is also visible in the plot of the source HR versus the intensity reported in Fig.~\ref{fig:HID}. 

\begin{figure}
\begin{center}
\includegraphics[angle=0,width=7.8cm]{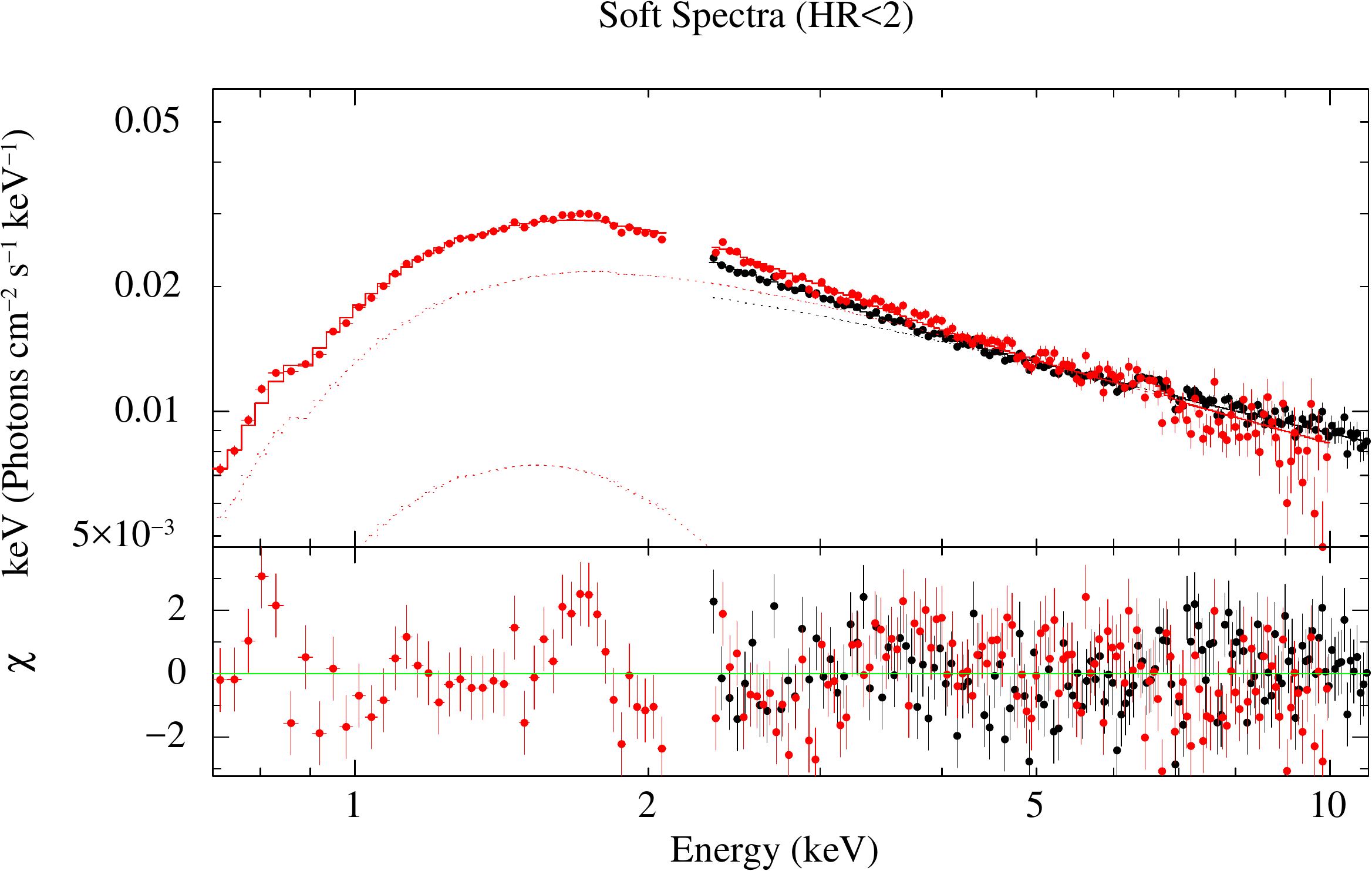}
\includegraphics[angle=0,width=7.8cm]{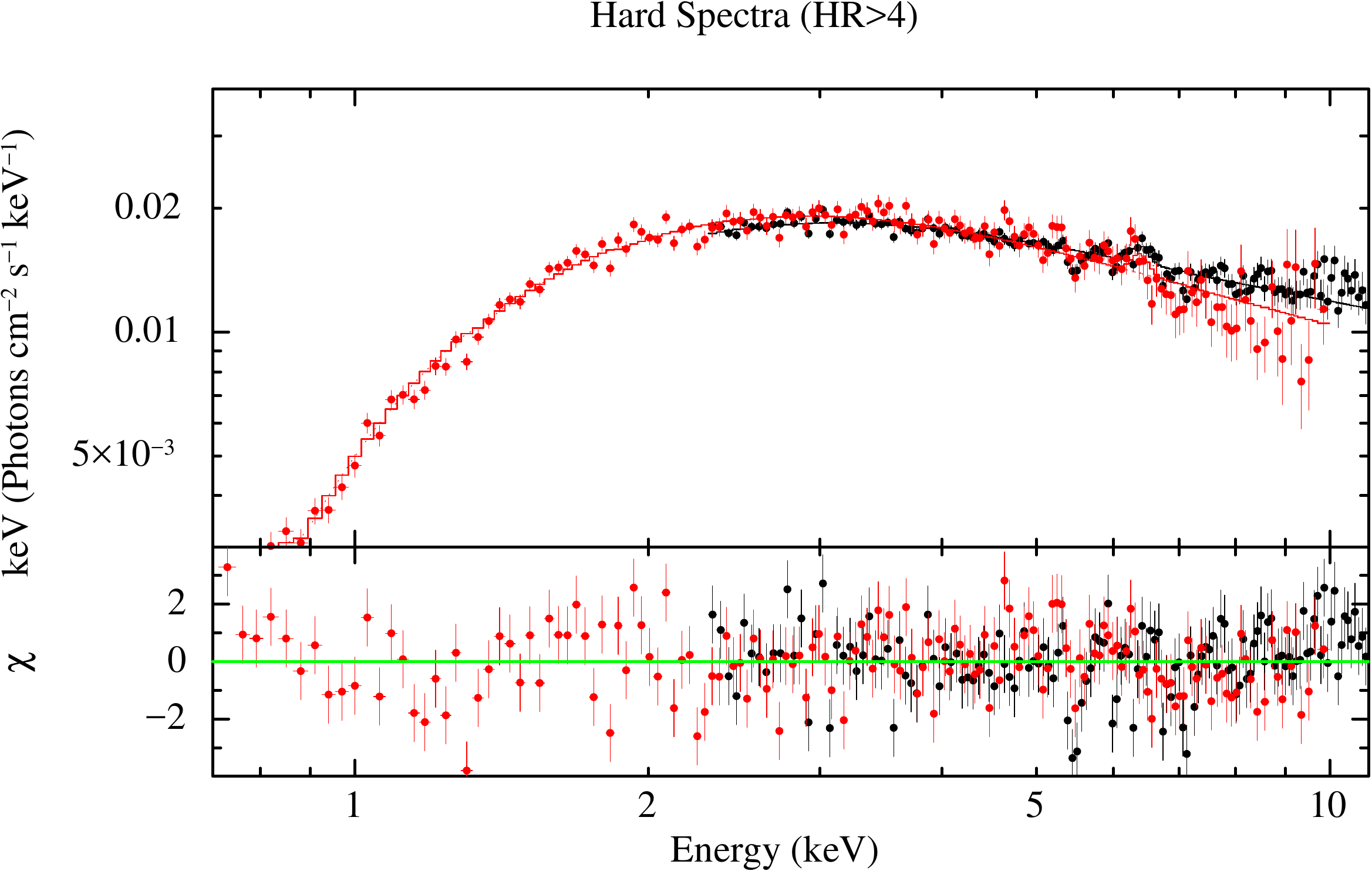}
\caption{Unfolded EPIC-pn (black) and MOS2 (red) spectra extracted separately during the time intervals corresponding to high (top) 
and low (bottom) values of the HR in Fig.~\ref{fig:lc}. In the case of the soft spectra, the best fit model is obtained with an 
absorbed black-body plus a \textsc{nthcomp} component and a Gaussian line. In the hard spectra, the black-body component was 
not required by the data. 
The different model components are also represented in the bottom and lower plots with dashed lines. The residuals from 
the best fits are shown in the lower panels. Note that the \pn data below 2.3\,keV have been discarded in order to avoid known 
calibration uncertainties at energies $\lesssim$1.7\,keV and the two instrumental features at 1.8\,keV and 2.2\,keV (see also S16). 
In the case of the {\mos}2 spectrum, we also discarded the interval 2.1--2.3\,keV due to the presence of a strong instrumental features.} 
\label{fig:spectracomb}
\end{center}
\end{figure}

As the HR variations are most likely produced by changes in the spectral parameters of the source, 
we extracted two sets of EPIC spectra by stacking together 
the time intervals corresponding to HR$<$2 and HR$>$4. These two HR values were chosen by visually inspecting the 
light curves in Fig.~\ref{fig:lc} and to maximize the variability between the two sets of spectra. 
The result is shown in Fig.~\ref{fig:spectracomb}. The softer spectra are characterized by a significantly enhanced emission 
below $\sim$2~keV. An acceptable fit to the soft spectra is obtained by using the same model adopted by S16 to describe 
the broad-band spectrum\footnote{We did not report here the analysis of the source average \xmm spectrum as this has been 
exhaustively discussed by S16.} of \igr, i.e. an absorbed black-body plus an \textsc{nthcomp} Comptonized component and a Gaussian 
to take into account the iron K$\alpha$ line emission (the complete model in \textsc{Xspec} is 
\textsc{Constant*TBabs*(BBodyrad+nthcomp+Gauss)}). No absorption column density variations can be evidenced between the two spectra.
In the hard spectrum, we did not find evidence for a significant 
\textsc{BBodyrad} component and thus we did not include it in the fit. We found that, if such component is included in the fit,
with a temperature fixed at that determined from the soft spectra, then the upper limit on the normalization would be a factor 
of 15 lower than that in the soft spectra. As reported by S16,  
we found a discrepancy between the spectral slope of the EPIC-pn and the MOS2 data of about 6\% that we attribute to  
remaining calibration uncertainties of the EPIC instruments in timing mode. 
Following their approach, we did not tie together in the fit the value of the photon index of the \textsc{nthcomp} component 
between the \pn and the {\mos}2 spectra. The results obtained by fitting the HR resolved spectra are summarized 
in Table~\ref{tab:spefits}. Note that we did not make use of the \rgs data because the 
read-out time of this instrument is too large to be able to follow the fast HR variations displayed by the source (typically from less 
than a second to a few seconds at the most). We verified that there are no significant changes in the pulse profile of the source and 
in the measured pulsed fraction during the time intervals corresponding to different HR with respect to the values reported in S16 
for the full observation. 
\begin{table}
\centering
\caption{Results of the fits to the combined \pn and \mos spectra, extracted for different hardness ratios. 
In the table $N_{\rm H}$ is the absorption column density, $kT_{\rm BB}$ ($R_{\rm BB}$) the temperature (radius) 
of the black-body component, $\Gamma$ the slope of the \textsc{nthcomp} component, and $kT_{\rm bb}$ 
($kT_{\rm e}$) the temperature of the soft Comptonized photons (temperature of the Comptonizing electrons). 
We also indicated with $E_{\rm k\alpha}$, $\sigma_{\rm k\alpha}$, and $\mathrm{EQW}_{\rm k\alpha}$ the energy, width, 
and equivalent width of the iron K$\alpha$ line. Uncertainties are expressed at 90\% confidence level. }
\label{tab:spefits}
\begin{tabular}{@{}llllllllllllll@{}}
\hline
Parameter & HR$>4$ & HR$<$2\\
\hline	
\smallskip		
$N_{\rm H}$ (10$^{22}$~cm$^{-2}$) & 0.35$\pm$0.02 & 0.31$^{+0.03}_{-0.05}$ \\
\smallskip
$kT_{\rm BB}$ (keV) & --- & 0.42$^{+0.02}_{-0.04}$ \\
\smallskip
$R_{\rm BB}$$^{a}$ (km) & --- & 3.8$^{+2.1}_{-0.4}$ \\
$\Gamma$ (pn) & 1.55$\pm$0.02  & 1.61$\pm$0.01 \\
$\Gamma$ (MOS2) & 1.68$\pm$0.05  & 1.69$\pm$0.02 \\
\smallskip
$kT_{\rm e}$$^{b}$ (keV) & 29  & 29 \\
\smallskip
$kT_{\rm bb}$ & 0.63$\pm$0.02 & 0.29$^{+0.20}_{-0.06}$ \\
\smallskip
$E_{\rm k\alpha}$ & 6.38$\pm$0.08 & 6.52$\pm$0.07\\
\smallskip
$\sigma_{\rm k\alpha}$ & 0.13$^{c}$ & 0.13$\pm$0.06\\ 
\smallskip
$\mathrm{EQW}_{\rm k\alpha}$ & 0.032$\pm$0.12 & 0.024$_{-0.09}^{+0.022}$ \\
\smallskip
$F_{\rm 0.5-2~keV}$$^{d}$ (10$^{-11}$~erg~cm$^{2}$~s$^{-1}$) & 2.0$^{+0.2}_{-0.1}$ & 5.0$^{+0.1}_{-0.2}$ & \\
\smallskip
$F_{\rm 2-10~keV}$$^{d}$  (10$^{-10}$~erg~cm$^{2}$~s$^{-1}$) & 1.90$\pm$0.03 & 1.73$\pm$0.02 & \\
$\chi^2_{\rm red}$/d.o.f. & 1.40/294 & 1.48/298 \\
$C_{\rm pn}$$^{e}$ & 1.0 & 1.0 \\
$C_{\rm MOS1}$$^{e}$ & 1.08$\pm$0.02  & 1.15$\pm$0.02 \\
Exp. Time (ks) & 6.0 & 23.6 \\
\hline \\
\end{tabular}
\begin{list}{}{} 
\scriptsize
\item[$^{\mathrm{a}}$:] This is evaluated for a distance of 4~kpc. 
\item[$^{\mathrm{b}}$:] This value could not be well determined in the fit and thus we fixed to the value measured from 
the source average spectrum reported by S16. 
\item[$^{\mathrm{c}}$:] This parameter could not be reliably measured from the fit and was thus fixed to 
the value determined from the fit to the HR$>$4 spectra.
\item[$^{\mathrm{d}}$:] This is the observed flux, i.e. not corrected for absorption.
\item[$^{\mathrm{e}}$:] We fixed to 1 the normalization constant of the EPIC-pn spectrum.  
\end{list} 
\end{table}

\section{The timing analysis}
\label{sec:timing}

In order to pursue the investigation on the mechanism triggering the atypical variability of \igr,\ we performed 
a dedicated timing analysis of the \pn and \nustar data focusing on the aperiodic noise rather than on the coherent pulsation, investigated by S16.  
We first converted the arrival time of each photon detected to the solar system barycenter using the known optical 
position of the source \citep{torres2008}.

\begin{figure}
\begin{center}
\includegraphics[angle=0,width=8.1cm]{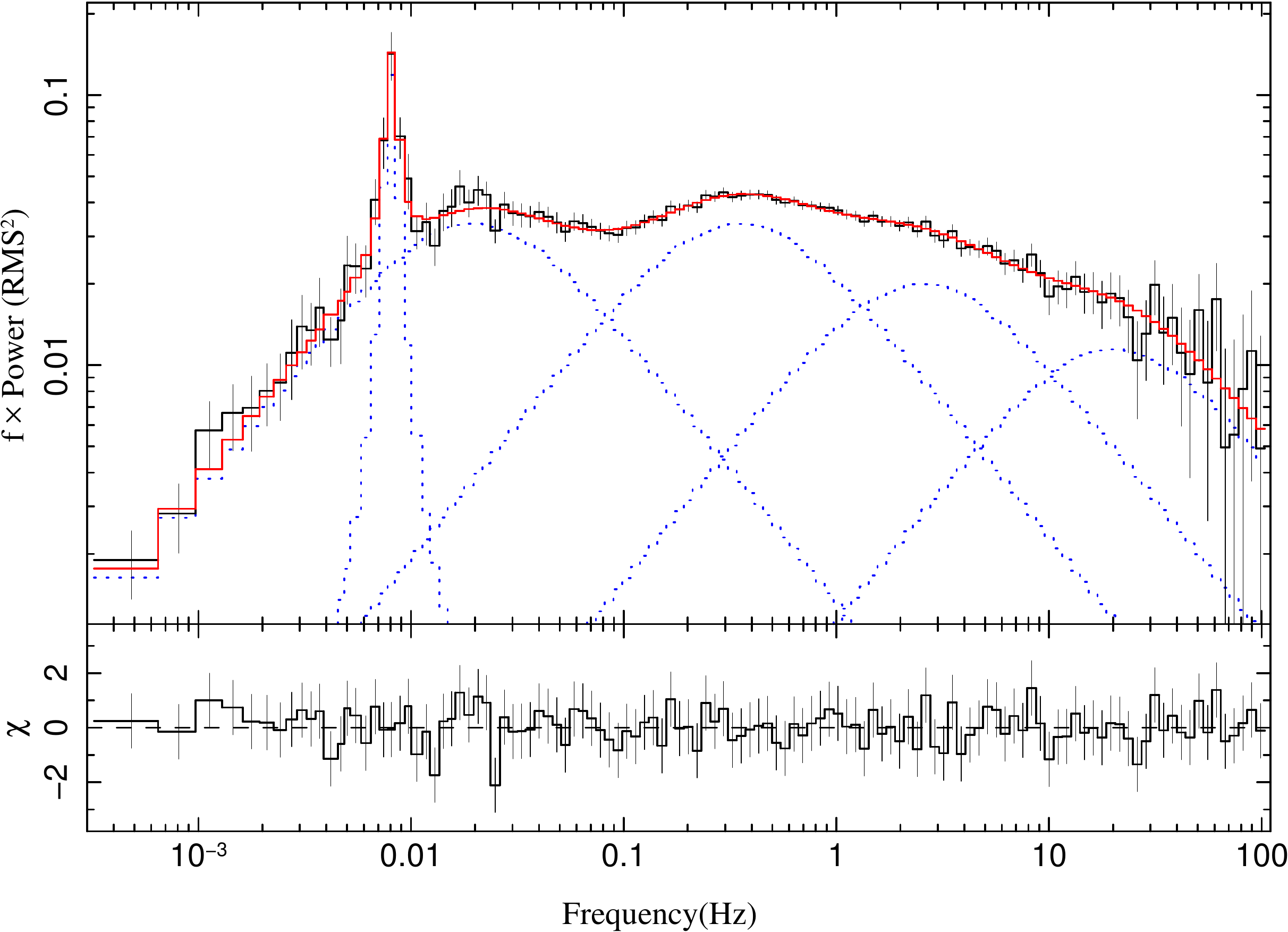}
\caption{Average power spectrum of \igr\ as measured by the \pn in the 0.5--11\,keV energy range. The best-fit model to the power-spectrum 
(red line in the upper panel) is obtained by using the Lorentzian functions indicated in Table~\ref{tab:psd} and represented as blue dotted lines. 
The residuals from the best fit 
are reported in the bottom panel. The prominent QPO at $\sim$8\,mHz is well visible in the plot.} 
\label{fig:psd}
\end{center}
\end{figure}

We extracted a power spectrum of the source in 
the entire energy range available (0.5--11\,keV) from the \pn light curve binned at 
2.952\,ms. The power spectrum was averaged over 23 intervals covering each $2^{20}$ bins (i.e., about 3095\,s) 
and then logarithmically rebinned with a factor of 0.1. From the fit to the power spectrum in the Lehay normalization, 
we measured a white noise of $1.993\pm0.002$ and subtracted the latter from the data before performing any other investigation. 
The source power spectrum in the rms$^2$ normalization \citep[see][for a comprehensive review on timing techniques]{Uttley2014} is shown 
in Fig.~\ref{fig:psd}. 

\begin{figure}
\begin{center}
\includegraphics[angle=0,width=8.1cm]{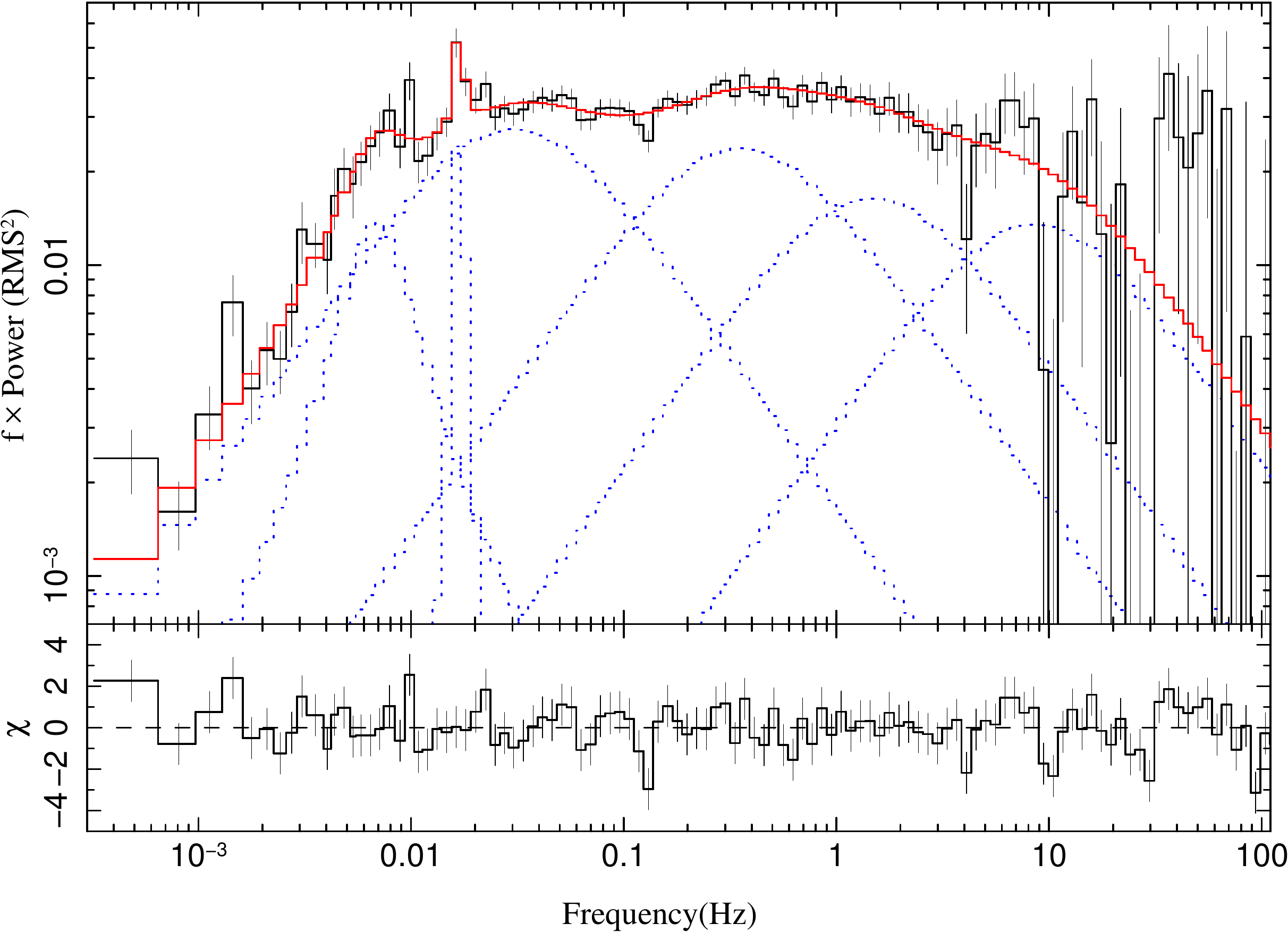}
\caption{Average \igr\ cospectrum as measured by the \nustar\ FPMA and FPMB in the 3--60\,keV energy range. 
The best-fit model (solid red line in the upper panel) is obtained by using the Lorentzian functions indicated in 
Table~\ref{tab:psd} and represented as blue dotted lines. The residuals from the best fit are reported in the bottom panel.} 
\label{fig:psd_nustar}
\end{center}
\end{figure}

For \nustar, we extracted the source light curves in the 3--60\,keV energy band for both the FPMA and FPMB detectors with bins of 2.952\,ms 
and computed the cospectrum, i.e. the real part of the cross power density spectrum,
as a best estimate of the white-noise subtracted power density spectra. This procedure is necessary to take into account the effect of 
dead time in the FPMA and FPMB as detailed in \citet{bachetti2015}. We computed the cospectrum in stretches of
$2^{20}$ bins, discarding intervals with gaps due to the satellite orbit around the Earth and other screening criteria. 
This procedure ensures a sample of 11 cospectra, whose average is then logarithmically rebinned  with a step of 0.1. The result 
is plotted in Fig~\ref{fig:psd_nustar}.

As commonly done for other AMXPs, we used a set of zero-centered Lorentzians to fit the 
broad-noise component of the power (cross) spectrum and included additional Lorentzian components to take into account the presence 
of frequency-confined features.
For the zero-centered Lorentzians, we used the expression:
\begin{equation}
\mathrm{L}_i(f) = \frac{\pi}{2} \frac{\mathrm{rms}_i^2 W_i}{W_i^2 + f^2}\,,
\label{eq:L}
\end{equation}
where $f$ is the frequency and $W_i$ is the half width of the Lorentizian. For the two Lorentzians describing the features at 8\,mHz and 16\,mHz, we used the expression: 
\begin{equation}
\mathrm{QPO}_i(f) = \frac{1}{\pi}\left[\frac{1}{2}-\frac{\tan^{-1} (-2Q_i)}{\pi}\right]^{-2} 
\frac{2\,\mathrm{rms}_i^2 Q_i \nu_i}{\nu_i^2 + 4Q_i^2 \left(f-\nu_i\right)^2}\,,
\label{eq:QPO}
\end{equation}
where ``QPO'' stands for Quasi Periodic Oscillation, $Q_i$ is the quality factor defined as the centroid 
frequency ($\nu_i$) of the feature divided by its full width at half maximum (FWHM).
This function can be used to describe broader noise components with characteristic frequency 
$\nu_i\left(1+\frac{1}{4Q_i^2}\right)^{1/2}$, when the quality factor is of order unity or lower. 
All timing analysis and fits were performed using ISIS \citep{isis} expanded by SITAR\footnote{\url{http://space.mit.edu/cxc/analysis/SITAR/}}.
The results of model fitting are computed
in the frequency range $3.2\times10^{-4}$--165\,Hz and
summarized in Table~\ref{tab:psd}.

\begin{table}
\caption{Best-fit parameters obtained from the fit to the average source \pn (\nustar) power (co) spectrum with a model comprising 
5 (6) Lorentzian components, two of which are used to take into account the presence of 
narrow frequency features at $\sim$8 and 16\,mHz (QPOs). Q indicates the quality factor of the 
QPO, while W is the width of the Lorentzian function (see Eq.~\ref{eq:L}--\ref{eq:QPO}). 
Uncertainties are expressed at 90\% confidence level.}
\begin{center}
 \begin{tabular}{l r@{}l r@{}l}
\hline
 Parameter & \multicolumn{2}{c}{\pn}&\multicolumn{2}{c}{\nustar}\\
 \hline
L$_1$ rms & 0.206 & $\pm0.006$  & 0.187 & $\pm0.007$\\
\smallskip
L$_1$ W [Hz] & 0.0198 & $\pm0.0018$ & 0.030 & $_{-0.005}^{+0.006}$\\
\smallskip
L$_2$ rms & 0.205 & $\pm0.005$  & 0.174 & $_{-0.036}^{+0.008}$\\
\smallskip
L$_2$ W [Hz] & 0.35 & $\pm0.03$ & 0.35 & $\pm0.09$\\
\smallskip
L$_3$ rms & 0.160 & $\pm0.008$ & 0.14 & $_{-0.05}^{+0.02}$\\
\smallskip
L$_3$ W [Hz] & 2.5 & $\pm0.5$  & 1.5 & $_{-0.7}^{+1.4}$\\
\smallskip
L$_4$ rms & 0.120 & $\pm0.012$  & 0.131 & $_{-0.046}^{+0.002}$\\
\smallskip
L$_4$ W [Hz] & 20 & $^{+11}_{-7}$ & 9 & $_{-4}^{+23}$\\
\smallskip
QPO$_1$ rms & 0.15 & $\pm0.02$  & 0.12 & $\pm0.02$\\
\smallskip
QPO$_1$ Q & 9 & $^{+5}_{-3}$ & 1.2 & $_{-0.4}^{+0.8}$\\
\smallskip
QPO$_1$ $\nu$ [Hz] & 0.00807 & $\pm0.00017$  & 0.0063 & $\pm0.0007$\\
\smallskip
QPO$_2$ rms & -- & -- & 0.066 & $\pm0.016$\\
\smallskip
QPO$_2$ Q & -- & --  & $>$ &6\\
\smallskip
QPO$_2$ $\nu$ [Hz] & -- & -- & 0.0168 & $_{-0.0005}^{+0.0003}$\\	
$\chi_\mathrm{red}^2$/d.o.f & 0.45 &/107 & 1.29&/99\\ 
\hline
\end{tabular}
\label{tab:psd}
\end{center}
\end{table}

In the \pn data, we find a prominent narrow feature at $\sim$8\,mHz, resembling a 
QPO \citep[see, e.g.,][for a recent review]{wang2016}, 
while the broad-band noise is characterized by four zero-centered Lorentzians (L$_1$--L$_4$). In the harder \nustar band, the broad-band noise
has a similar shape, while the narrow 8\,mHz feature disappears,
substituted by a narrow weak feature at roughly double its frequency (QPO$_2$ in Table~\ref{tab:psd} with rms 7\%) and by a broader 
component at a slightly lower frequency (6\,mHz). 

In order to investigate further the strong QPO at 8\,mHz, visible only in the \xmm data, 
we divided the EPIC-pn energy range in 12 intervals characterized by the same number of 
detected X-ray photons. For each interval, we extracted a light curve binned at 0.2952\,s.
We computed the power density spectra averaging over intervals of $2^{11}$ time bins (i.e., about 604\,s resulting in 119 samples). 
This choice ensures that the QPO remains coherent 
in each time interval (based on the lower boundary of the uncertainty we determined on the QPO quality factor in 
Table~\ref{tab:psd}) and all its power is confined in a single Fourier bin.
Then, we
rebinned them geometrically with a factor of 0.1, and normalized them to the usual rms$^2/f$ convention.
All rebinned and normalized power spectra are modeled by three zero-centered Lorentzians with width $\sim$0.02, $\sim$0.4\,Hz, and 2.5\,Hz  
(the latter fixed at the value of Table~\ref{tab:psd}), plus 
two narrower Lorentzians to take into account the QPO at 8\,mHz and a second feature around 16\,mHz. 

\begin{figure}
\begin{center}
\includegraphics[angle=270,width=8cm]{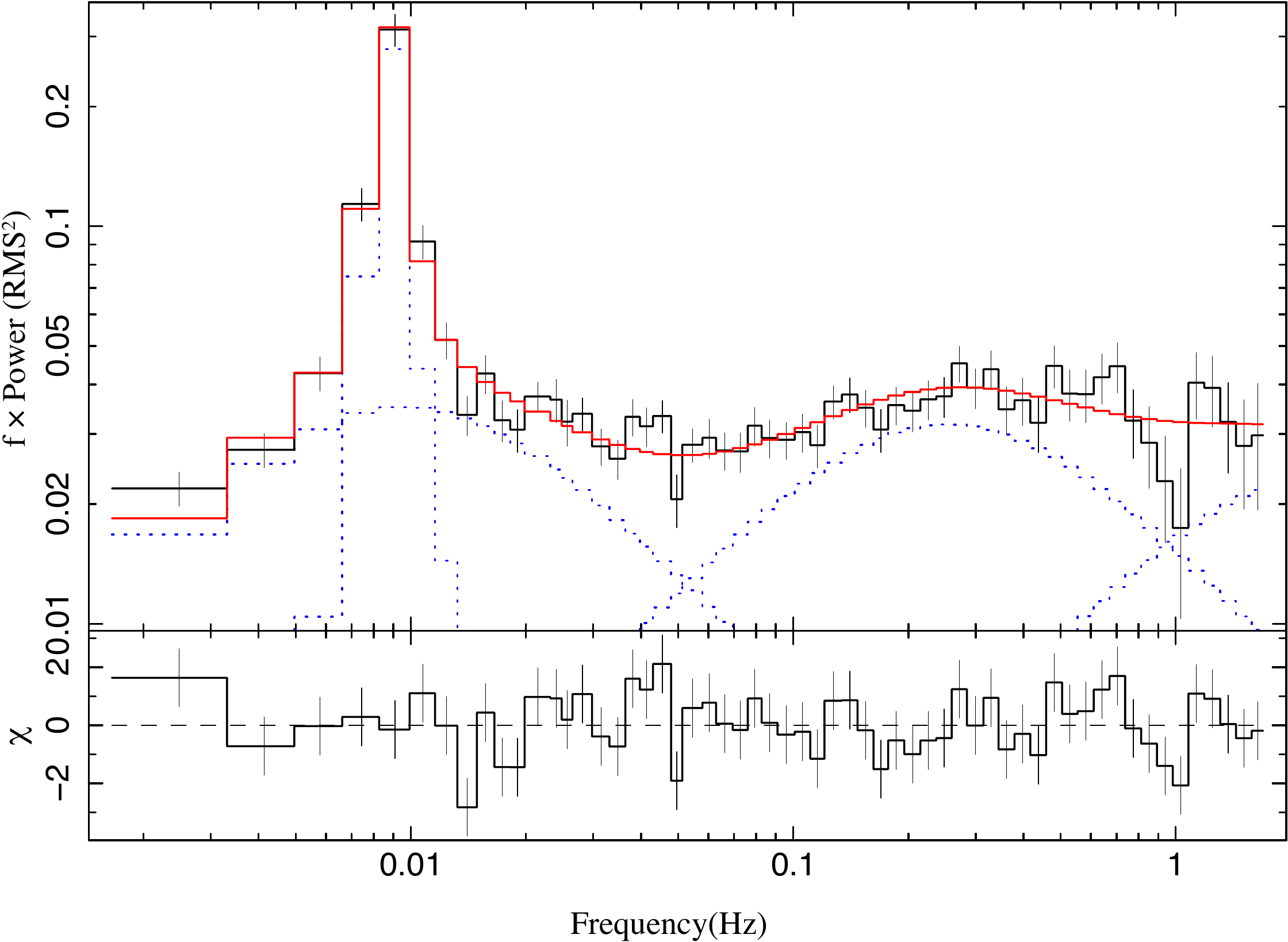}
\includegraphics[angle=270,width=8cm]{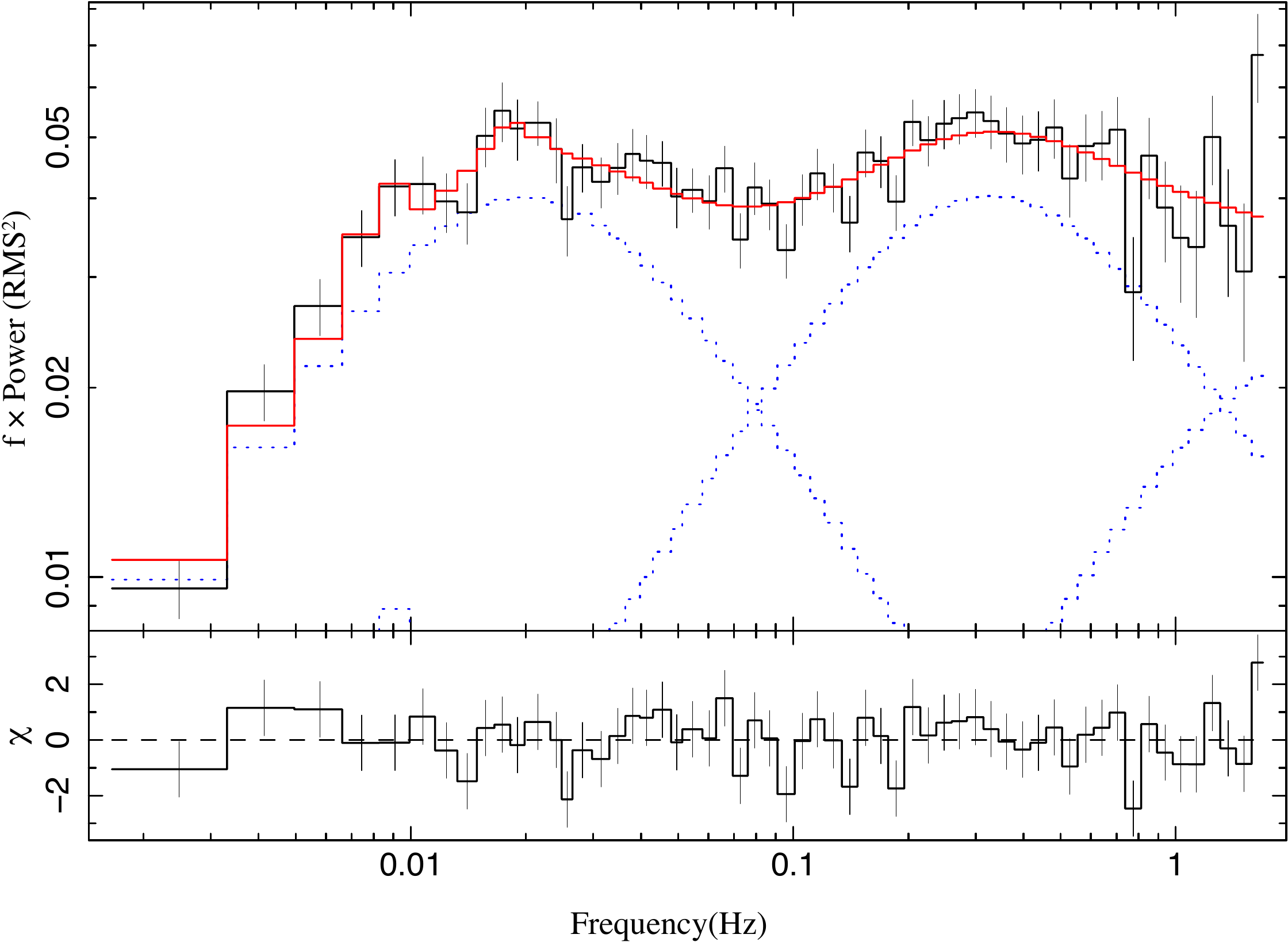}
\caption{Two examples of average power spectra obtained from the EPIC-pn data in two energy bands: 0.94--1.16 keV (top) and 
4.5--6.0\,keV (bottom). The best fit models are indicated in both cases with a red solid line, the single Lorentzian components as blue dotted lines,  while the residuals from the fits are shown in the bottom panels.}
\label{fig:crosspec}
\end{center}
\end{figure}

Two examples of \xmm power spectra are shown in Fig.~\ref{fig:crosspec}, together with the corresponding best-fit models and the residuals 
from the fits.  

\begin{figure}
\begin{center}
\includegraphics[angle=0,width=2.8cm]{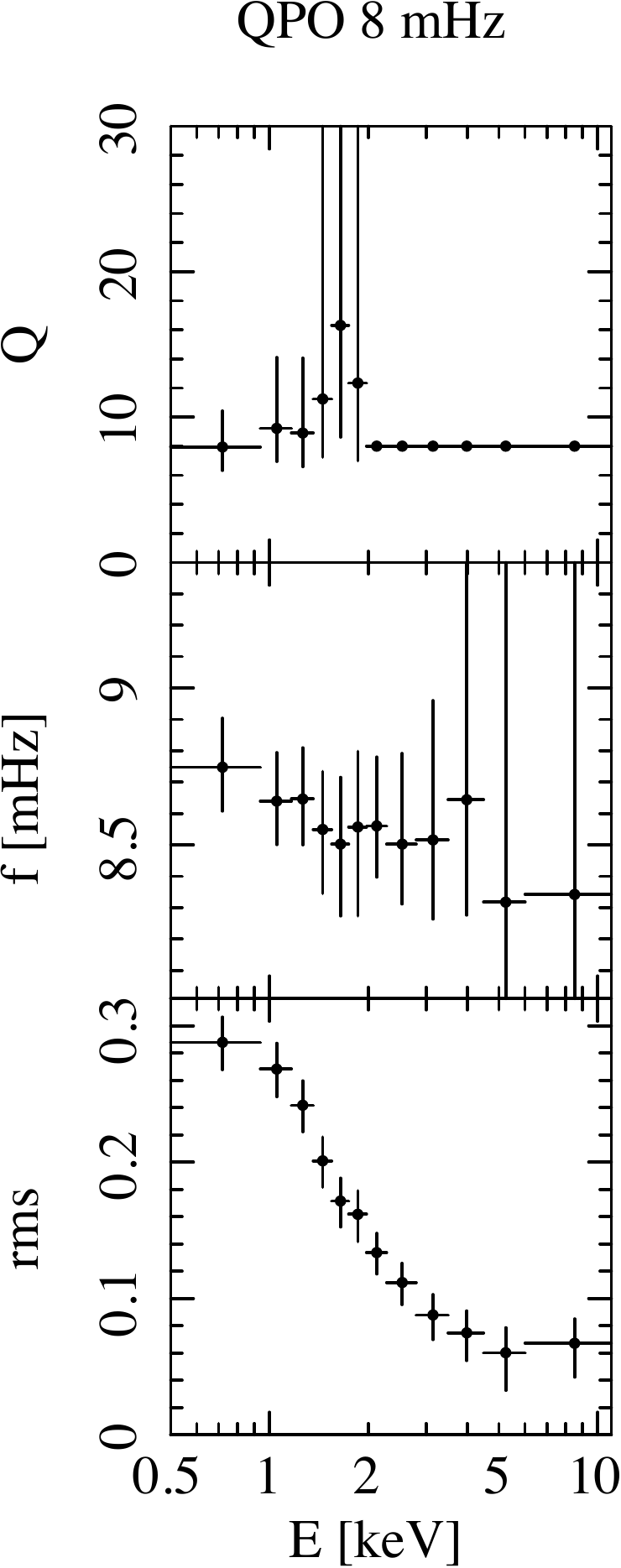}\hfill
\includegraphics[angle=0,width=2.8cm]{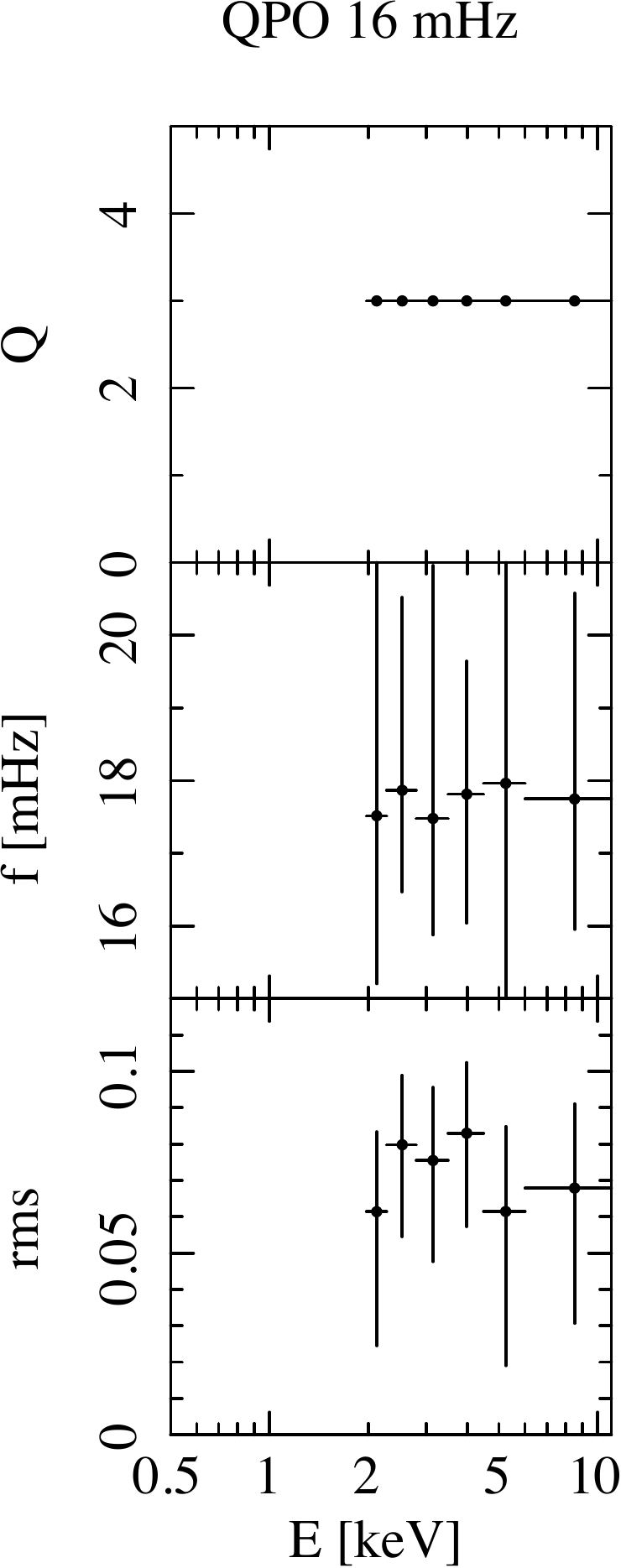}\hfill
\includegraphics[angle=0,width=2.8cm]{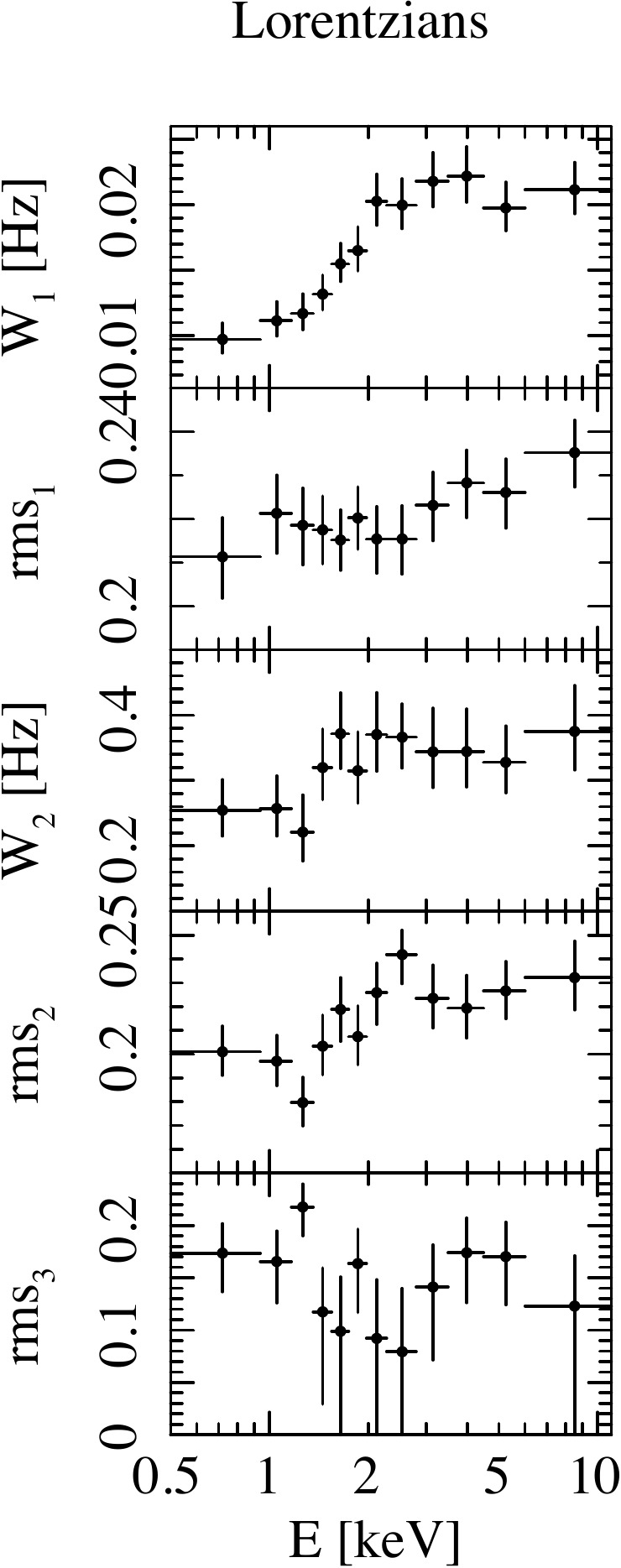}
\caption{Energy dependence of the parameters of the QPO at 8~mHz (left), the feature at 16~mHz (middle), and the three 
broad-band noise Lorentzian components (right) detected in the \xmm\ data of \igr.\ Note that in several energy bins the value 
of $Q$ for the QPOs could not be reliably determined from the fit and its value was fixed (points without error bars). 
The feature at 16~mHz is not significantly detected in the cross-spectra 
extracted at energies $\lesssim$2~keV. The third Lorentzian has a fixed width of 2.5\,Hz. Uncertainties 
are represented at 90\% confidence level.}  
\label{fig:rms}
\end{center}
\end{figure}

In Fig.~\ref{fig:rms}, we show the energy dependency of the best-fit parameters measured from the \pn data for the two features 
at 8\,mHz and 16\,mHz, together with those of the broad-noise Lorentzians. The frequency of the QPO at 8\,mHz remains constant at all energies, 
while its rms dramatically decreases above $\sim$1.5--2\,keV. The quality factor Q, 
is poorly determined owing to our choice of frequency binning. We have verified that the 
trend on rms would be maintained with other reasonable choices of binning. 
The feature at 16\,mHz disappears at those energies at which the QPO at 8\,mHz is 
more prominent and seems to be significantly broader than this other feature. To check whether the higher frequency feature 
is the harmonic of the QPO at 8~mHz, we applied the method outlined in \citet{Ingram2015}. For each light curve in one 
of the defined energy band for \xmm that was Fourier transformed, we computed the ratio between the modules of the Fourier 
transform and the phase difference at the respective frequencies. We found that the computed values of the ratios and phase differences do not cluster 
around any preferred values, at odds with what is expected if the two features at 8\,mHz and 16\,mHz are linked. 
We thus suggest that these feature are not correlated and could be possibly produced by different mechanisms. 

\begin{figure}
\begin{center}
\includegraphics[angle=0,width=7.8cm]{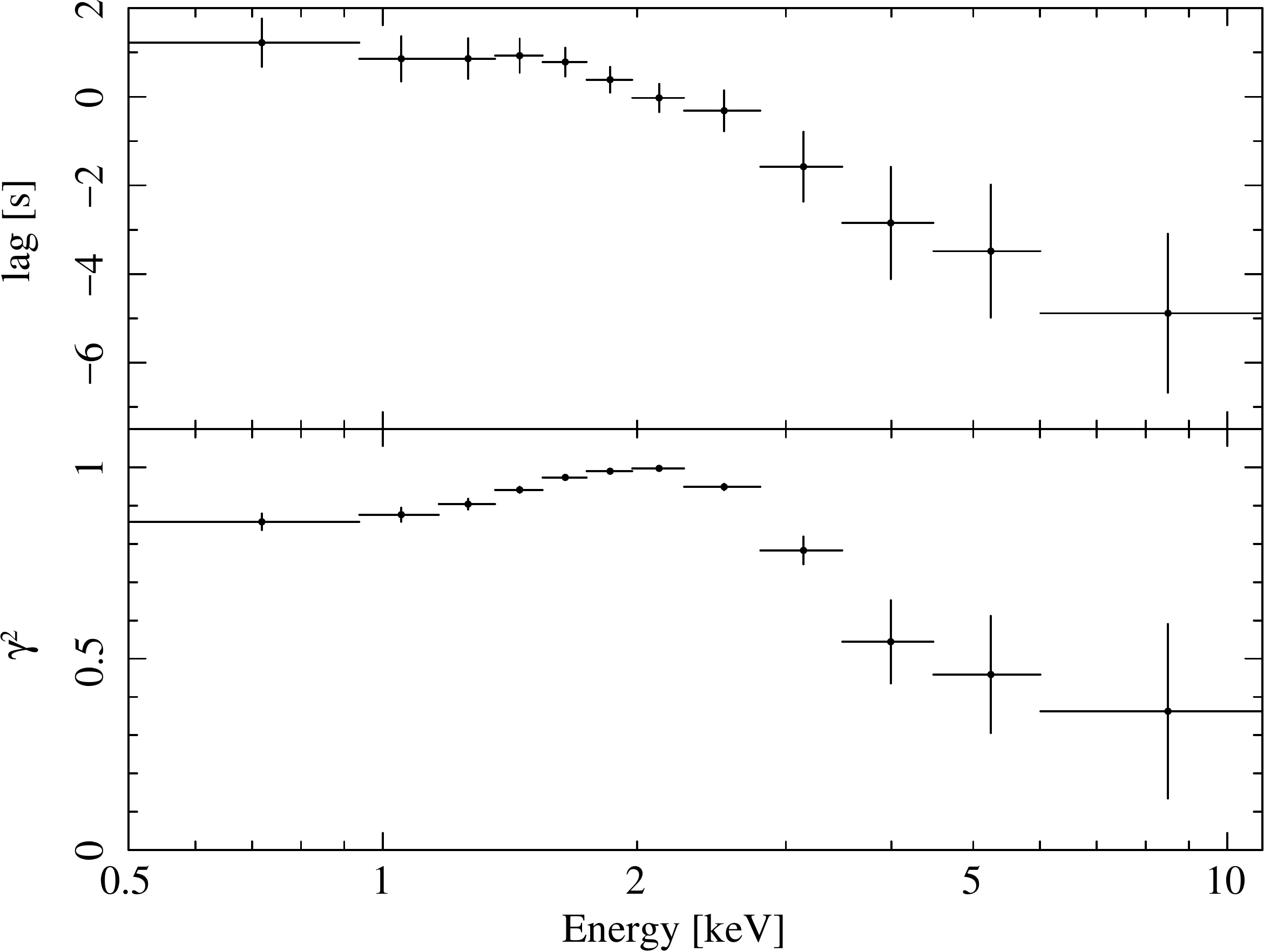}
\caption{{\it Upper panel}: time lags at the QPO frequency (8\,mHz) computed in each energy bin with respect to the reference energy band 
(i.e., the entire EPIC-pn energy band excluding each time the energy bin where the lag is calculated). 
{\it Lower panel}: the corresponding cross-correlation linear coefficient. Uncertainties are given at 68\% confidence level.}
\label{fig:lags}
\end{center}
\end{figure}

The prominent QPO at 8\,mHz detected in both the averaged and energy-resolved \xmm power 
spectra of \igr\ is clearly associated with the X-ray variability 
of the source that, as shown in Sect.~\ref{sec:spectra}, undergoes continuous flares. The typical 
time scale of the rise and decay of these flares is comparable with the QPO frequency. By inspecting
the source dynamical power spectra, we also verified that the 8\,mHz QPO does not drift in frequency and it is 
present throughout the entire observation.
As the intensity variations of the source during the flares are also accompanied by significant spectral changes 
(see Sect.~\ref{sec:spectra}), we investigated the dependence of the energy distribution of the source 
X-ray emission as a function of the QPO phase. We follow the procedure outlined by \citet{Ingram2015} and 
\citet{Ingram2016}.  A key ingredient in this method is the average phase lag of the
source signal at the QPO frequency, compared to a reference band.
This can be obtained in each energy band by averaging the complex argument 
of the cross spectra over the 119 intervals of 2$^{11}$ bins defined before. The cross spectrum is computed 
assuming as band of interest the energy intervals used above and as reference band the full \pn energy band minus 
the energy band of interest. To obtain the uncertainty on the phase lag (Eq.~12 in \citealt{Uttley2014}), 
we computed the average coherence and its uncertainty using Eq.~(2) and (8) in \citet{Vaughan1997}. 
In Fig.~\ref{fig:lags}, we show the resulting correlation  and the time lags at the frequency of the QPO. 
It can be noticed that as long as the QPO is characterized by a fractional rms larger 
than 0.1 at energy $\lesssim$3.5\,keV, the light curve at the QPO frequency is linearly correlated to the average \pn signal, whereas
the linear correlation is reduced at higher energies.
This analysis reveals that the modulation of the X-ray flux at the frequency of the QPO has a delay of few seconds 
on the soft X-ray photons compared to those at the higher energies ($\gtrsim$1.5-2~keV).  

\begin{figure}
\begin{center}
\includegraphics[angle=0,width=8cm]{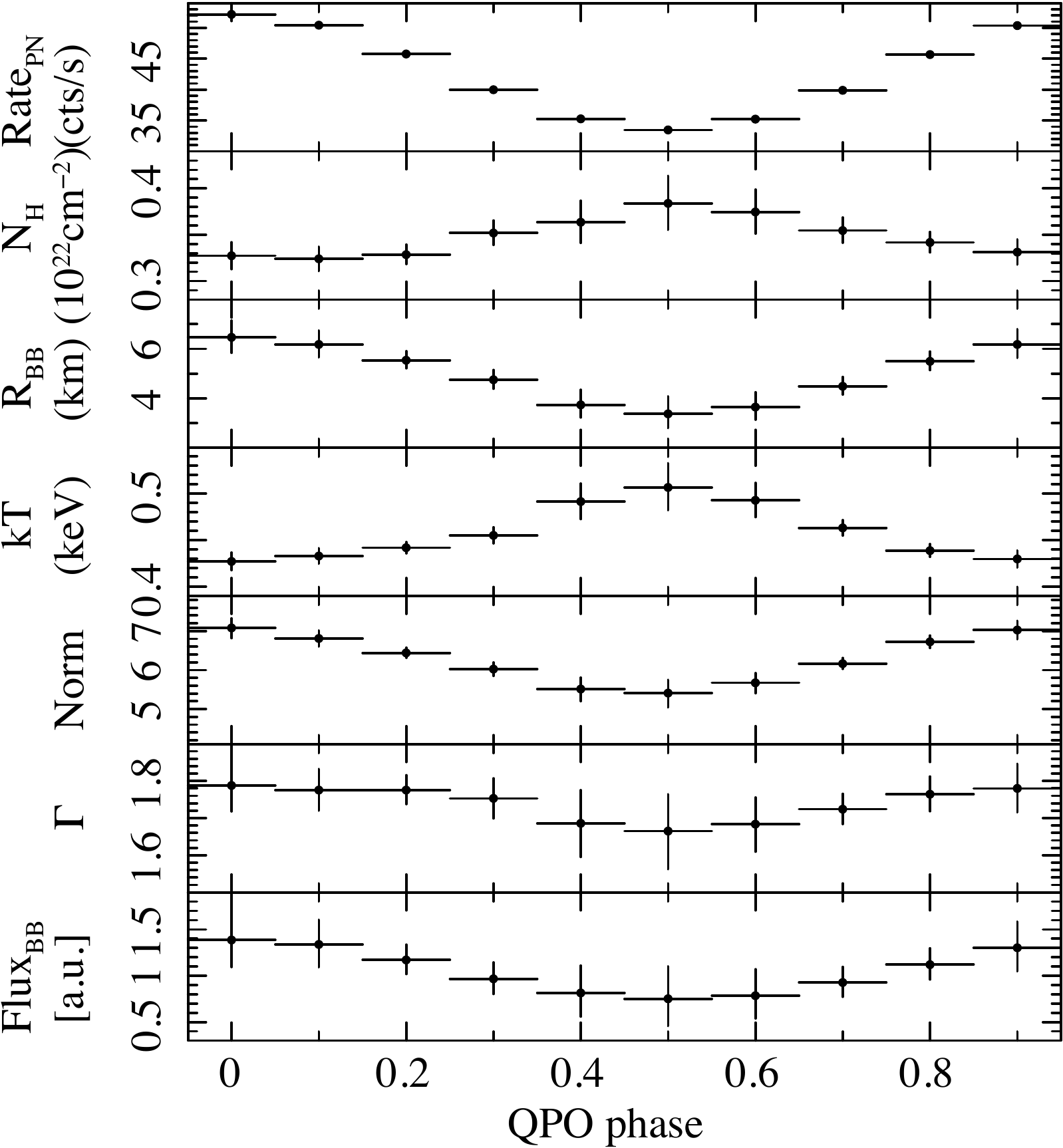}
\caption{Top panel: the derived \pn count rate as function of the QPO phase. 
Other panels: best-fit parameters obtained from the fits to the QPO phase-resolved spectra of \igr.\  
The model includes absorbed Comptonization (\textsc{nthcomp}) and black-body components. 
The phase zero corresponds to the maximum of the QPO. $\Gamma$ and $Norm$ indicates the spectral slope and 
the normalization of the \textsc{nthcomp} component, the latter represented in units of $10^{-3}$. The flux of the black body 
(lowermost panel) is not a parameter of the fit and is estimated as $r_\mathrm{BB}^2 kT_\mathrm{BB}^4$ and expressed in arbitrary units (a.u.).
Uncertainties are expressed at a 90\% confidence level.} 
\label{fig:qpo_phase}
\end{center}
\end{figure}

The energy-dependent wave front of the QPO in our 12 energy bands can 
be described in the Fourier space by:
\begin{equation}
W_j(E) =\mu(E) \sigma_j(E) e^{-i\Phi_j(E)}\,
\end{equation}
where $\mu(E)$ is the mean count-rate of the source, while $\sigma_j(E)$ and $\Phi_j(E)$ are the fractional rms and the phase 
shift of the $j$-th harmonic, respectively. In the case of \igr, only the fundamental of the QPO was detected and thus $j=1$. The parameter 
$\sigma_1(E)$ can be obtained from the fit of the real part of the average cross spectrum (the values are reported in Fig.~\ref{fig:rms}), 
while the phase shift $\Phi_1(E)$ computation is described in the previous paragraph.
A set of source spectra accumulated at different QPO phases ($\phi$) can be extracted by transforming the QPO wave front 
in the time domain using the following expression: 
\begin{equation}
w(E,\phi) = \mu(E)\left(1+\sqrt{2} \sigma_1 \cos{\left[ \phi -\Phi_1(E)\right]}\right)\,.
\end{equation}
Here $\mu(E)$ can be obtained by 
appropriately rebinning the average \pn and \mos2 spectra. 
Owing to the lower effective area of the \mos2, we use the timing results from the \pn for both instruments.
To derive the uncertainties on $w(E,\phi)$, we adopted a bootstrapping technique: 
10\,000 wave fronts were computed for each energy bin by varying their rms and phase shift within the corresponding 
confidence intervals assuming a Gaussian distribution. The resulting mean value of the distribution was then 
assumed as the best estimate of $w(E,\phi)$, while the standard deviation of the distribution is considered as the 
associated 1~$\sigma$ uncertainty. The obtained spectra were then fit with the same model\footnote{We did not include the 
iron line due to the reduced energy resolution of these spectra compared to those discussed in Sect.~\ref{sec:spectra}.} 
used for the HR resolved spectra: 
\textsc{Constant*TBabs*(BBodyrad+nthcomp)}. As for the spectral analysis of Sect.~\ref{sec:spectra}, we 
used different slopes for the power-law for the two instruments and an inter-calibration constant. 
We fixed in the fit the value of $kT_{\rm e}$=29~keV as done in Sect.~\ref{sec:spectra} and $kT_{\rm bb}$=0.9\,keV, according 
to the results found by S16 (none of these two parameters could be reliably constrained). 
Despite the limited energy resolution and the uncertainty in the QPO front reconstruction, it was possible to extract up to 
10 spectra at different QPO phases, still detecting significant variations on the corresponding spectral 
parameters. The results of this analysis are summarized in Fig.~\ref{fig:qpo_phase}, where the black-body 
radius is calculated by assuming a distance of 4\,kpc (DF16) and the parameter $\Gamma$ is the one 
determined for the \pn.\ The \mos2 parameters follow a similar trend with slightly different absolute values and
larger uncertainties, while the cross-calibration term was verified to be constant and then fixed to its average value.
The plot suggests a moderate variability of spectral slope and absorption column density
over the QPO phases. The radius and temperature of the black-body, and the \textsc{nthcomp} normalization
display the most noticeable variations. We also derived the flux of the black-body, which shows how the variations of temperature and 
radius compensate each other at a certain extent, showing that the thermal emission is responsible for only a part of the flux variability.

\begin{figure}
\begin{center}
\includegraphics[angle=0,width=8cm]{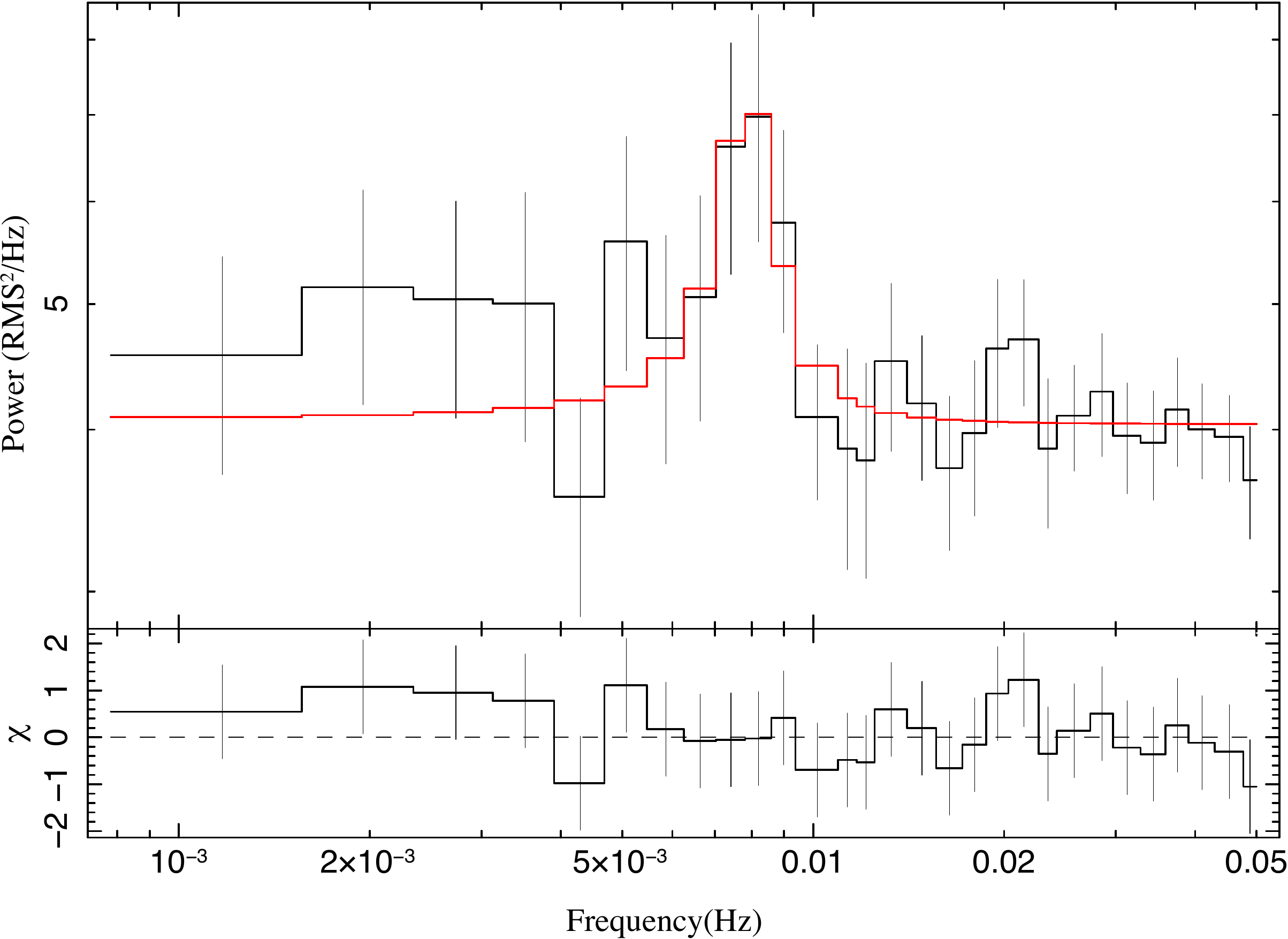}
\caption{Power density spectrum of \igr\ derived from the time series of the fractional pulsed amplitude of the 
pulsed signal extracted in intervals of ten seconds. The red curve represented the best fit model to the data. 
The corresponding residuals are also shown.} 
\label{fig:pulsed_psd}
\end{center}
\end{figure}

We also investigated if and how the QPO affects the pulsed emission recorded from \igr.\ Using the ephemeris of S16,
we extracted, every ten seconds during the \xmm\ observation, a pulse profile of the source with 16 phase bins 
in the energy range 0.5--3\,keV, where 
the QPO is more prominent. Then, we derived the value of the fractional amplitude in each case by measuring 
the amplitude of the fundamental mode of the pulsed emission and dividing it by the source average intensity 
during the considered time interval. The power density spectrum of the time series of the 
fractional amplitudes could then be obtained with a reasonably good resolution in frequency by using 
stretches of 64 bins (we also applied a logarithmic rebinning of 1.1). The resulting 
54 independent power spectra were then averaged assuming in each case a 100\% uncertainty to obtain the 
results shown in Fig.~\ref{fig:pulsed_psd}. This averaged PSD was normalised in accordance to the rms$^2$ convention 
and then fit with a constant plus a QPO function. We measured a normalization constant of 4.0$\pm$0.2 and obtained for the 
QPO parameters: $\nu=7.9\pm0.7$\,mHz, $Q=5_{-2}^{+13}$, and norm=$(9_{-3}^{+2})\times10^{-2}$ at 90\% c.l. 
As the averaged PSD was not obtained from a Poissonian signal the parameters derived from the fit loose their usual 
meaning, but the detection of the QPO unambiguously indicates that the mechanism giving rise to the aperiodic 
variability in \igr\ is also affecting the amplitude of the source pulsed emission.  

\section{Discussion and conclusions}
\label{sec:discussion}

In this paper we investigated the origin of the peculiar flaring-like X-ray variability showed by \igr\ during its 
latest outburst in 2015 and giving rise to a prominent 8~mHz QPO in the source power spectra extracted at energies $\lesssim$2~keV. 
This behavior is not common among AMXPs in outburst, but 
there are a few examples of both neutron stars and black-hole binaries in which an X-ray 
variability on a similar time scale was detected. 

A flaring like-behavior in the mHz band, which vaguely resemble what we observed from \igr, was recorded in 2010 from the transient 
neutron star LMXB IGR\,J17480$-$2446 during its first detectable outburst \citep{bordas10}. 
In this occasion, the thermonuclear X-ray bursts from the source were observed to increase in frequency and decrease in intensity 
while the source was approaching the peak of the outburst. When the persistent luminosity of the sources reached 10-50\% of the Eddington 
value, the bursts were so regularly spaced one from the other to produce a prominent QPO in the power spectrum of the source at about 2-5\,mHz. 
The main difference with the case of \igr is that the ``flares'' observed around the peak of the outburst from IGR\,J17480$-$2446 
were clearly of thermonuclear origin and the corresponding spectral energy distribution could be very well described by assuming 
a single black-body component \citep[see, e.g.][and references therein]{linares2012}. 

Our HR-resolved spectral analysis carried out in Sect.~\ref{sec:spectra} reveal that the flares observed from \igr\ are typically 
(but not always) characterized by a softer emission close to the peak and a much harder emission during the decay. 
However, in none of these cases, the spectral energy distribution could be successfully described with a simple thermal component, 
being the \textsc{nthcomp} component the dominant contribution in all observable energy range. 
The softer spectra are characterized by a prominent thermal component that 
is not present in the harder spectra (or at least largely suppressed). The Comptonization component does not display comparable 
large changes, as confirmed by the lack of any major variation in the 2-10~keV X-ray flux compared to the 
factor $\sim$4 recorded in the 0.5-2~keV energy band. According to the usual scenario of mass accreting onto a NS in a LMXB, 
it seems reasonable to assume that the thermal component is most likely generated onto the surface of the compact object, 
as the temperature is around 0.3-0.6~keV and the radius is of only a few km. The Comptonization processes giving rise to the 
harder spectral component can occur in the accretion column of the NS or closer to the surface of the disk or in both locations 
\citep{falanga07}. Thus, the spectral variability accompanying the flares of \igr\ is more complex than that expected 
during thermonuclear bursts. This disfavors the possibility that the flares from \igr\ have a thermonuclear origin. 

The AMXP \sax\ is known to display a flaring-like behavior both close to the peak of the outburst (the so-called ``high 
luminosity flaring'') and late in the outburst decay (``low luminosity flaring''). During these periods, the source shows 
flares with typical rise and decay times of 0.2--1~s, giving rise to prominent QPOs at frequencies of 1--5~Hz 
\citep[see][and references therein]{patruno09,bult14}. These features have been interpreted in terms of the so-called 
``dead-disk instability'', according to which accretion can proceed in an unstable way when the inner disk boundary is 
located close to the co-rotation radius where the propeller mechanism is supposed to set-in 
\citep[see, e.g.,][and references therein]{romanova2004}. 
In this scenario, modest variations of the instantaneous accretion rate can move the inner disk radius inside and outside co-rotation, 
alternating between short periods of enhanced and inhibited accretion \citep[see, e.g.,][]{angelo}. The precise time scale on which the 
inner boundary of the disk is moved inward or away from the compact object (and thus the correspondingly generated QPO frequency) 
depends mainly on the physical conditions in the accretion disks and on the properties of the NS (spin period and magnetic field strength). 
As \sax\ and \igr\ are located at a similar distance and both achieve a typical peak X-ray flux during outbursts of 
few$\times$10$^{-9}$\,erg\,cm$^{-2}$\,s$^{-1}$  (see Fig.~1 in DF16), we should expect to observe in both systems a QPO produced by 
this mechanism at roughly the same frequency (note that the spin period of \igr\ is shorter than that of \sax, but the co-rotation 
radius is only 30\% larger). This conclusion challenges the applicability of the dead disk model to the case of \igr,\ as the frequency of 
the QPO reported in Sect.~\ref{sec:timing} is $\gtrsim$100 times smaller than that of the QPOs detected from \sax.\ Furthermore, 
the dependence of the QPO rms on energy is opposite in the two sources: the rms of the QPOs observed from \sax\ 
dramatically increases with energy while the QPO reported in \igr\ is only detectable below $\sim$2~keV. This suggests that the 
two phenomena are most likely triggered by different mechanisms. A similar conclusion applies to the case of the AMXP 
NGC~6440~X-2. This source also showed during its outburst in 2010 a flaring-like behavior and a prominent QPO at 1~Hz with  
an rms increasing strongly with energy \citep{patruno10,patruno13}.

Some dipping sources also displayed a strong quasi periodic variability with a typical frequency of 1\,Hz and an high rms amplitude
over a band-limited noise \citep{Homan2012}. These QPOs have been interpreted as being due to self-obscuration of the precessing 
accreting material, which is misaligned from the rotational axis of the pulsar. At odds with these sources, however, 
\igr is not characterized by a high inclination (X-ray eclipses are not detected) and its QPO has  
a completely different frequency and energy dependence (disappearing above $\sim$2~keV instead of being virtually 
achromatic as in dipping sources).

Two more striking cases are those of 4U~1636$-$536 and 4U~1608$-$52, two X-ray bursters 
and Atoll sources which displayed in a few occasions soft X-ray QPOs ($\lesssim$5~keV) with frequencies of about 
8\,mHz and rms of a few percents \citep{Mike2001,Altamirano2008}. These oscillations were detected just before the occurrence of 
a type-I burst in a specific limited range of the source X-ray flux and observed to disappear immediately afterwards. 
In a few observations, the QPO also reappeared several kiloseconds after the burst \citep{Lyu2015,Lyu2016}. 
A similar phenomenon was also reported in the case of the intermittent AMXP Aql~X-1. The disappearance of the QPO immediately after the 
onset of a thermonuclear explosion led to the suggestion that this feature could be related to 
episodes of marginally stable nuclear burning. As explained theoretically by \citet{Heger2007}, in this situation 
the X-ray luminosity of the source can display oscillations on the time scale of roughly two minutes but the expected 
range of mass accretion rates for which such a phenomenon could manifest is a factor of $\sim$10 higher than that 
measured from 4U~1636$-$536 and 4U~1608$-$52 when the QPOs were observed. A possible way out could be to assume that 
the nuclear burning only occurs on a small fraction of the NS surface, even though this would require magnetic field intensities for the 
compact object much larger ($\sim10^{10}$\,G) than those expected in these systems (10$^{8-9}$\,G).

In the case of \igr, the possibility that the QPO is associated with the marginally stable nuclear burning  
is very hard to investigate as the sole and only 
type-I X-ray burst ever detected from the source occurred roughly three days before the \xmm and \nustar\ observations (see DF16). 
As we cannot firmly conclude about any association between the QPO and the nuclear burning, we investigate below the 
alternative possibility \citep[mentioned also by][]{Mike2001} 
that the variability observed from \igr\ is driven by a mechanism more similar to that operating in 
black-hole binaries (BHBs) and giving rise to the so-called ``heartbeats'' \citep[see, e.g.,][]{belloni2000,altamirano11}.
This phenomenon produces a flare-like variability on times scales similar to those observed from \igr\ and the spectral changes measured 
during the peaks and valleys of these flares have been associated with the movement of the inner disk boundary closer or further away 
from the event horizon. Even though the variability shown by the BHBs during the heartbeats is rather complicated, 
it is generally found that at the peak of the flares the temperature (radius) of the thermal component increases (decreases) 
and the opposite occurs during minima \citep{neilsen11,capitanio11,mineo12}. The origin of this behavior is not fully understood but it could 
be related to the Lightman-Eardley instability of the accretion disk that appears when the central source approaches the Eddington luminosity \citep{done2004}. 
While this interpretation would work well in the case of the bright BHB GRS\,1915+105, its application to the case of the sole other BHB 
displaying heartbeats, IGR\,J17091$-$3624, is more challenging \citep[but see also the cases of 
H1743$-$322 and IC~10~X-1][]{Altamirano2012,Pasham2013}. 
The distance to the latter system and the mass of the BH hosted in it are 
not known, but the luminosity observed from this source during the heartbeat state is believed to be orders of magnitude lower than that 
of GRS\,1915+105, unless the system is located beyond 20~kpc or the central BH is endowed with the lowest mass ever measured for such an object 
\citep{altamirano11}. Therefore, it cannot be excluded that heartbeats can also appear in low luminosity systems as \igr.

The luminosity achieved by \igr during the \xmm observation is of the order of $\sim$10$^{36}$\,erg\,s$^{-1}$ 
(assuming a distance of 4\,kpc), and thus far from the Eddington limit. 
The spectral changes measured during the flares is also dissimilar to what is observed from the BHBs displaying heartbeat variability, 
but we should remember that in the case of the present source there is a NS at the center of the accretion disk and thus it would be reasonable 
to expect different properties for the X-ray emission. 
According to the interpretation of a heartbeat variability, we can speculate that the suppression of the thermal component 
during the states of high HR (occurring most of the times close to the minima of the flares - that should correspond also to phase 
0.5 of the QPO) is due to the fact that when the disk is further away from the compact object the size of the hot spot on the 
neutron star surface decreases \citep[see, e.g.,][]{frank02}. 
If the displacement of the disk leads to a lower mass accretion rate, in the most simplistic accretion scenario we would expect the 
temperature of the thermal emission to decrease as well. Our findings in Fig.~\ref{fig:qpo_phase} suggest that the temperature 
has an opposite trend as function of the QPO phase, compared to what expected (i.e., an increase rather than a decrease around the QPO phase 0.5). 
However, we cannot exclude that in a more realistic accretion scenario the temperature responds in a less trivial 
way to changes in the inner boundary of the disk. 
If the thermal emission undergoes a strong Comptonization close to the NS surface, the plasma could be
cooled less effectively above a smaller hotspot, thus inverting the trend of the temperature as a function of the QPO phase. 
Investigating this point in more details would require theoretical improvements to the spectral 
models, which are beyond the scope of this paper. 
Apart from the black-body temperature, the remaining results obtained from the QPO phase 
resolved spectral analysis match well with the overall heartbeat interpretation. We found a larger radius for the thermal component 
at the maximum of the QPO, accompanied by a significant change in the normalization (but not in the slope) of the 
Comptonization component.

The results obtained from the timing analysis presented in Sect.~\ref{sec:timing} are also compatible with the heartbeat interpretation. 
The analysis of the fractional pulsed amplitude time series confirmed that the intensity of the pulsed emission, 
related to the extension of the hot spot(s) on the NS surface, is also modulated at the same QPO frequency. This supports the idea that 
the hot spot changes shape on a compatible time scale.
The measured time lags at the QPO frequency (Fig.~\ref{fig:lags}) are nearly constant below $\sim$2\,keV, where the QPO coherence and rms 
are high, and show a large negative shift above this energy. As these time lags are of the order of a few seconds and cannot be associated 
to any reasonable light travel time within a system as compact as a LMXB, it is likely that they are associated   
to geometrical effects due to changes in the region(s) where different emission components are produced. 

We note that the outburst of 2015 is not the first one in which quasi-periodic timing variability of \igr was pointed 
out. As discussed by \citet{linares07}, the source displayed already during the outburst in 2004 a particularly rich variety of timing 
features in the 3--25~keV energy range (as observed by the RXTE/PCA). The properties of the source broad-band noise between the 2015 and 
2004 outburst are qualitatively similar, although the component at 2.5\,Hz was characterized by a significantly higher rms amplitude 
and the narrow component at 0.16~Hz was not detected. A number of QPO discovered in the RXTE data with frequencies comprised 
in the 10--100\,mHz range led \citet{linares07} to conclude that \igr\ was showing an X-ray variability much more reminiscent of that of a 
BHB rather than of a NS LMXB. Even though the frequencies of the previously reported QPOs are not too dissimilar from the feature at 
8~mHz reported in the present paper, they could be very well detected in the higher energy 
bound covered by the PCA. This is at odds with our findings in Sect.~\ref{sec:timing}, which proved that the QPO at 8~mHz reported here for the 
first time is a purely soft phenomenon and becomes undetectable at energies $\gtrsim$2\,keV. It seems thus correct to consider different 
origins for all these QPOs. Note that the RXTE/PCA does not cover the energy range where the \xmm\ QPO was detected, and thus we cannot rule out 
that this feature was also present during the 2004 outburst. The much less significant feature we found at 16~mHz in the \xmm power spectra 
extracted at energies above 2~keV and in the \nustar cross-spectrum 
could more closely resemble one of the features reported before by \citet{linares07}. However, it is much less pronounced and in 
our case the statistics available in the hard spectral band is far too low to carry out a more complete comparison with the previous 
RXTE results.

In conclusion, we suggest that the X-ray variability and the associated mHz QPO discovered for the first time 
during the outburst of \igr in 2015 are likely produced by an heartbeat-like mechanism. The possibility that this 
variability is associated with phases of quais-stable nuclear burning cannot either be excluded or more solidly tested,
due to the paucity of type-I X-ray bursts observed from the source since its earliest detection in 1998.

\section*{Acknowledgements}
We thank the anonymous referee and Phil Uttley for their precious suggestions.
This work is based on observations obtained with \xmm  (OBSID 0744840201) and \nustar (OBSID~90101010002); 
the former is an ESA science mission with instruments and 
contributions directly funded by ESA Member States and the USA (NASA); the latter is 
a project led by the California Institute of Technology, 
managed by the Jet Propulsion Laboratory, and funded by the National Aeronautics and Space Administration. 
This research has made use of the \nustar Data Analysis Software (\textsc{NuSTARDAS}) 
jointly developed by the ASI Science Data Center (ASDC, Italy) and the California Institute of Technology (USA).
We also exploited the ISIS functions 
provided by ECAP/Remeis observatory and MIT (\url{http://www.sternwarte.uni-erlangen.de/isis/}).

\bibliographystyle{mnras}
\bibliography{igr}

\bsp	
\label{lastpage}
\end{document}